\documentclass[aps,pra,preprint,groupedaddress,amsmath,amssymb,showpacs]{revtex4-1}
\usepackage{amssymb}
\usepackage{booktabs}
\usepackage{graphicx,amsmath,amssymb,epsfig,color}
\usepackage{dcolumn}
\usepackage{bm}
\begin{document}
\title{Theoretical analysis on quantum interference effect in fast-light media}
\author{Datang Xu$^{1}$, Chaohua Tan$^{1}$,
        and Guoxiang Huang$^{1,2, }$\footnote{gxhuang@phy.ecnu.edu.cn}}
\affiliation{$^{1}$State Key Laboratory of Precision Spectroscopy and Department of Physics,
                   East China Normal University, Shanghai 200062, China\\
             $^{2}$NYU-ECNU Joint Physics Research Institute at NYU-Shanghai, Shanghai 200062, China
            }
\date{\today}

\begin{abstract}
We make a systematic theoretical analysis on the quantum
interference (QI) effects in various fast-light media (including
gain-assisted $N$, gain-assisted ladder-I, and gain-assisted
ladder-II atomic systems). We show that such
fast-light media are capable of not only completely eliminating the
absorption but also suppressing the gain of signal field, and hence
provide the possibility to realize a stable propagation of the
signal field with a superluminal velocity. We find that there is a
destructive (constructive) QI effect in gain-assisted
ladder-I (gain-assisted N) system, but no QI in the
gain-assisted ladder-II system; furthermore, a crossover from
destructive (constructive) QI to Autler-Townes splitting may occur for
the gain-assisted ladder-I (gain-assisted N) system when the control
field of the system is modulated. Our theoretical
analysis can be applied to other multi-level systems, and the
results obtained may have promising applications in optical and
quantum information processing and transmission.
\end{abstract}
\maketitle

\section{INTRODUCTION}\label{Sec:1}

In the past two decades, much attention has been paid to the study of slow light~\cite{Milo05},
which can be realized in various optical media~\cite{Khurgin2009}. The most typical system for
obtaining slow light is the use of electromagnetically induced transparency (EIT) occurring in a
three level $\Lambda$-type atomic system interacting with two resonant laser
fields~\cite{Fleischhauer2005}. Slow light has many practical applications, including
high-capacity communication networks, ultrafast all-optical information processing, precision
spectroscopy and precision measurements, quantum computing and quantum information, and so
on~\cite{Milo05,Khurgin2009,Fleischhauer2005}.

However, as pointed out in Ref.~\cite{Deng1}, EIT-based slow-light scheme has some drawbacks.
Two of them are significant signal-field attenuation and spreading
and very long response time. Parallel to the study of slow light,
in resent years there are also tremendous interest
on the investigation of fast light (also called superluminal light)~\cite{note00,Boyd1,Boyd2}.
Chu and Wong~\cite{Chu} firstly demonstrated a superluminal propagation of optical
wave packet in an absorptive medium. In order to
suppress the substantial attenuation of the optical wave packet occurred in the experiment~\cite{Chu},
Chiao~\cite{Chiao} proposed to use a gain medium with inverted atomic
population to obtain a stable superluminal propagation.
Steinberg and Chiao~\cite{Steinberg} proved that the stable superluminal propagation
in the medium with a gain doublet is indeed possible. The works carried out by Wang
{\it et al}.~\cite{Wang} and Biglow {\it et al.}~\cite{Bigelow}, as well as those
reported in Refs.~\cite{Deng1,Agarwal2004,Jiang1,Jiang2,Huang1,Hang1,Zhu1,Li,Zhu2,Deng2,Tan}, further revealed many intriguing
aspects of fast light, including the possibility of obtaining giant Kerr nonlinearity
and superluminal optical solitons. Recently, it has been demonstrated that the use
of fast-light media can realize quantum phase gates~\cite{Hang2} and light and quantum
memory~\cite{Lezama,Lett}.

It is well known that the physical mechanism of EIT is the quantum interference (QI) effect
contributed by control field, by which the absorption of signal field
can be greatly suppressed, i.e. an EIT transparency window is opened
in the absorption spectrum of the signal field. Furthermore, the QI also results in a drastic
change of dispersion and hence a large reduction of the
group velocity of the signal field~\cite{Milo05,Fleischhauer2005,Khurgin2009}. In addition,
it has been discovered recently that in such systems there exists an interesting crossover from EIT to Autler-Townes
splitting (ATS)~\cite{Agarwal1997,Abi-Salloum2010,Anisimov2011,
Giner2013,Tan2013,Zhu2013,Tan2014,Peng2014,Ding,Lu,Davuluri}.
It is natural to ask the question: Is it possible to have similar phenomena for fast-light media?

In this article, we give a definite answer to this question by
investigating the absorption spectra of several typical fast-light
media, including gain-assisted $N$ (GAN), gain-assisted ladder-I
(GAL-I), and gain-assisted ladder-II (GAL-II) atomic systems (Fig.~1). We
carry out systematic theoretical analyses and give clear physical explanations
on the QI effects occurring in these fast-light media
by extending the spectrum-decomposition method (SDM) developed recently for the EIT-ATS
crossover of slow light~\cite{Agarwal1997,Abi-Salloum2010,Anisimov2011,
Giner2013,Tan2013,Zhu2013,Tan2014,Peng2014,Ding,Lu,Davuluri}.
We
show that such fast-light media are capable of not only completely
eliminating the absorption but also suppressing the gain of signalmodulate
field, and hence provide the possibility to realize a stable  long-distance
propagation of the signal field with a superluminal velocity. We
find that there is a destructive (constructive) QI effect in the GAL-I
(GAN) system, but no QI in the GAL-II system; furthermore, a
crossover from destructive (constructive) QI to ATS may occur for the
GAL-I (GAN) system if the control field of the system is modulated.
Our theoretical analysis can be applied to
other gain-assisted multi-level systems (e.g. quantum dots, rare-earth ions in crystals, etc.),
and the results obtained may have promising applications in optical
and quantum information processing and transmission.

The remainder of the article is organized as follows. In Sec.~\ref{Sec:2}, we present
the model and analyze the QI effect in the GAN
system. In Sec.~\ref{Sec:3} and Sec.~\ref{Sec:4}, we carry out similar analyses and
provide related results for the GAL-I and GAL-II systems, respectively. Finally, in
Sec.~\ref{Sec:5} we give a discussion and a summary of the main results
obtained in this work.

\section{Quantum interference characters of the GAN system}\label{Sec:2}

\subsection{Model and linear dispersion relation}

\begin{figure}
\includegraphics[scale=0.78]{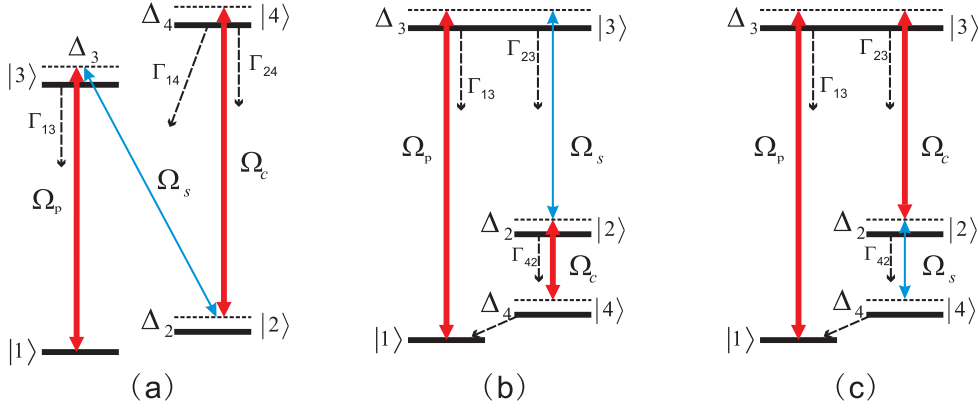}\\
\caption{(Color online) Excitation schemes of  four-state atoms with
the level configuration of the form (a) GAN system, (b) GAL-I
system, and (c) GAL-II system. In each system, four quantum states
$|1\rangle$, $|2\rangle$, $|3\rangle$, and $|4\rangle$ are coupled
by a pump field  ${\bf E}_p$ (with half Rabi frequency $\Omega_p$), a signal
field ${\bf E}_s$ (with half Rabi frequency $\Omega_s$), and a control field
 ${\bf E}_c$ (with half Rabi frequency $\Omega_c$) in different ways. $\Delta_j$
($j=2,3,4$) are detunings. The dashed arrows represent the
spontaneous emission or incoherent population exchange rates
(i.e. $\Gamma_{13}$, $\Gamma_{14}$,
$\Gamma_{24}$, $\Gamma_{23}$, $\Gamma_{42}$).
}\label{fig1}
\end{figure}

We first consider a cold atomic system with the GAN-type
level configuration (Fig.~\ref{fig1}(a)\,).
%
The quantum states $|1\rangle$ and $|3\rangle$ are coupled by a strong, continuous-wave (CW) pump
field (with half Rabi frequency $\Omega_p$), $|2\rangle$ and $|3\rangle$ are coupled by a weak,
pulsed signal field (with half Rabi frequency $\Omega_s$), and $|2\rangle$ and $|4\rangle$ are
coupled by a strong, CW control field (with half Rabi frequency $\Omega_c$), respectively.
$\Delta_j$ ($j=2,3,4$) are detunings, and  dashed arrows denote the spontaneous-emission rates
($\Gamma_{13}$, $\Gamma_{14}$, $\Gamma_{24}$). Such system, where the states $|j\rangle$
($j=1,2,3$), the pump field ($\Omega_p$) and signal field ($\Omega_s$) constitutes a core of
active Raman gain (ARG)~\cite{Jiang1,Jiang2,Huang1}, has attracted considerable attention
recently~\cite{Deng1,Agarwal2004,Hang1,Li,Tan}.

We assume the pump, signal, and control fields propagate in $z$ direction, the electric field
vector acting in the system reads $\mathbf{E}={\bf E}_p+{\bf E}_s+{\bf
E}_c=\sum_{l=p,s,c}\mathbf{e}_l{\cal E}_l {\rm exp}[i( k_l z-\omega_lt)]+$c.c., where
$\mathbf{e}_l$  $(k_l)$ is the unit polarization vector (wavenumber) of the electric-field
component with the envelope ${\cal E}_l$  $ (l=p,s,c)$. Under electric-dipole and rotating-wave
approximations, the interaction Hamiltonian of the system interacting with laser fields reads
$\hat{\cal H}_{\rm int}=-\hbar\left[\sum_{j=1}^4 \Delta_j|j\rangle\langle
j|+(\Omega_p|3\rangle\langle1|+\Omega_s|3\rangle\langle2|+\Omega_{c}|4\rangle\langle2|+{\rm
h.c.})\right]$, where h.c. represents Hermitian conjugation,
$\Omega_p=(\boldsymbol{p}_{31}\cdot{\cal E}_{p})/\hbar$, $\Omega_s=(\boldsymbol{p}_{32}\cdot{\cal
E}_s)/\hbar$, and $\Omega_{c}=(\boldsymbol{p}_{42}\cdot{\cal E}_{c})/\hbar$ are respectively the
half Rabi frequencies of the pump, signal, and control fields, with $\boldsymbol{p}_{jl}$ being
the electric-dipole matrix element associated with the transition from state $|j\rangle$ to state
$|l\rangle$.

Under electric-dipole approximation (EDA) and rotating-wave approximation (RWA),
the dynamics of the system is  governed by the Bloch equation~\cite{Boyd}
\begin{equation}\label{BEGAN}
i\hbar \left(\frac{\partial}{\partial
t}+\hat{\Gamma}\right)\sigma=\left[\hat{\cal H}_{\rm int},\sigma\right],
\end{equation}
where $\sigma$  is a $4\times 4$ density matrix in the interaction picture,
and $\hat{\Gamma}$ is a $4\times 4$ relaxation matrix describing the spontaneous
emission and dephasing. The explicit expression of Eq.~(\ref{BEGAN})
is presented in Appendix~\ref{appA1}.

As in Ref.~\cite{Deng1}, we assume
the one-photon detuning $\Delta_3$ is much larger than all the Rabi
frequencies, Doppler broadened line width (resulted by the thermal motion of the atoms),
atomic coherence decay rates, and frequency shift induced by the pump and control fields. In this
situation, the population keeps mainly in the ground state
$|1\rangle$ to guarantee the system working in fast-light regime; furthermore,
Doppler effect can be largely suppressed. However, the (remanent) Doppler effect still has
influence on the QI property of the GAN and GAL-II systems (but not in the GAL-I system),
as shown in Sec.~\ref{Sec:5}.

From the Maxwell equation $\nabla^2 {\bf E}_s-(1/c^2)\partial^2{\bf
E}_s/\partial t^2$=$(1/\epsilon_0c^2)\partial^2 {\bf P}_s/\partial t^2$
with the electric polarization intensity
${\bf P}_s= {\cal N}_a(\boldsymbol{p}_{23}\sigma_{32}\,
e^{i(k_s z-\omega_{s} t)}+{\rm c.c.})$,
one can obtain the equation of motion for the Rabi frequency of the signal field
under slowly-varying envelope approximation (SVEA), which reads
\begin{equation}\label{Maxwell-N}
i\left(\frac{\partial}{\partial
z}+\frac{1}{c}\frac{\partial}{\partial t}\right)\Omega_s
+\kappa_{23}\sigma_{32}=0,
\end{equation}
where $\kappa_{23}={\cal
N}_a\omega_s|\boldsymbol{p}_{32}|^2/(2\hbar\varepsilon_0 c)$ with
${\cal N}_a$ the atomic density. Note that in deriving the above
equation we have assumed the signal-field envelope is wide enough
in the transverse (i.e. $x$, $y$) directions, so that the diffraction
term $(\partial^2/\partial x^2+\partial^2/\partial y^2)\Omega_s$
can be disregarded.

The base state of the system (i.e. the steady-state solution of the Maxwell-Bloch (MB)
Eqs.~(\ref{BEGAN}) and (\ref{Maxwell-N}) for  $\Omega_s=0$) is
$\sigma_{11}^{(0)}=\Gamma_{14}|\Omega_{c}|^2(\Gamma_3X_{31}+|\Omega_{p}|^2)/D$,
$\sigma_{22}^{(0)}=\Gamma_{23}|\Omega_{p}|^2(\Gamma_4X_{42}+|\Omega_{c}|^2)/D$,
$\sigma_{33}^{(0)}=\Gamma_{14}|\Omega_{c}|^2|\Omega_{p}|^2/D$,
$\sigma_{44}^{(0)}=\Gamma_{23}|\Omega_{c}|^2|\Omega_{p}|^2/D$,
$\sigma_{31}^{(0)}=-\Omega_{p}\Gamma_{14}\Gamma_3X_{31}|\Omega_{c}|^2/(d_{31}D)$,
$\sigma_{42}^{(0)}=-\Omega_{c}\Gamma_{23}\Gamma_4X_{42}|\Omega_{p}|^2/(d_{42}D)$,
with $X_{31}=|d_{31}|^2/(2\gamma_{31})$,
$X_{42}=|d_{42}|^2/(2\gamma_{42})$ and
$D=\Gamma_{14}|\Omega_{c}|^2(\Gamma_3X_{31}+2|\Omega_{p}|^2)+\Gamma_{23}
|\Omega_{p}|^2(\Gamma_4X_{42}+2|\Omega_{c}|^2)$.
Here the meaning of the quantities $\Gamma_{jl}$ and $d_{jl}$ has been explained in the
Appendix~\ref{appA1}.  For large $\Delta_3$  one has
$\sigma_{11}^{(0)}\approx 1$, $\sigma_{31}^{(0)}\approx -\Omega_{p}/d_{31}$,
and all other $\sigma_{jl}^{(0)}\approx 0$, which means that
initially the atomic medium is prepared with the population mainly in the ground state state $|1\rangle$.

The base state of this system  will evolve into a
time-dependent state when the weak signal field is switched on~\cite{note000}. Solving the MB
Eqs.~(\ref{BEGAN}) and (\ref{Maxwell-N}) we obtain the solution
\begin{subequations}\label{First-N}
\begin{eqnarray}
&&\Omega_s=F\,e^{i\theta},\\
&&\sigma_{32}^{(1)}=\frac{B(\sigma_{33}^{(0)}-\sigma_{22}^{(0)})-
(D_{p}+|\Omega_{c}|^2)\Omega_{p}\sigma_{31}^{*(0)}-(D_{c}
+|\Omega_{p}|^2)\Omega_{c}\sigma_{42}^{*(0)}}{(\omega+d_{32})B-
|\Omega_{p}|^2(D_{p}+|\Omega_{c}|^2)-|\Omega_{c}|^2(D_{c}+|\Omega_{p}|^2)}F\,e^{i\theta},
\end{eqnarray}
\end{subequations}
where $F$ is an envelope (its concrete form is not needed here),
$\theta=K(\omega)z-\omega t$~\cite{note1},
$D_{p}=(\omega-d_{41}^*)(\omega-d_{43}^*)-|\Omega_{p}|^2$,
$D_{c}=(\omega-d_{21}^*)(\omega-d_{41}^*)-|\Omega_{c}|^2$,
and $B=(\omega-d_{21}^*)(\omega-d_{41}^*)(\omega-d_{43}^*)
-(\omega-d_{21}^*)|\Omega_{p}|^2-(\omega-d_{43}^*)|\Omega_{c}|^2$.
Explicit expressions of the other first-order solutions for $\sigma_{jl}^{(1)}$ ($j\neq 3, l\neq 2$)
are omitted here.

The linear dispersion relation of the system reads
\begin{equation}\label{K-N}
K(\omega)=\frac{\omega}{c}+\kappa_{23}\frac{B(\sigma_{33}^{(0)}
-\sigma_{22}^{(0)})-(D_{p}+|\Omega_{c}|^2)\Omega_{p}\sigma_{31}^{*(0)}-(D_{c}
+|\Omega_{p}|^2)\Omega_{c}\sigma_{42}^{*(0)}}{(\omega+d_{32})B
-|\Omega_{p}|^2(D_{p}+|\Omega_{c}|^2)-|\Omega_{c}|^2(D_{c}+|\Omega_{p}|^2)}.
\end{equation}
The group velocity of the signal field is given by
$V_{g}=[\partial \text{Re}(K)/\partial \omega]^{-1}$.

\begin{figure}
\includegraphics[scale=1.36]{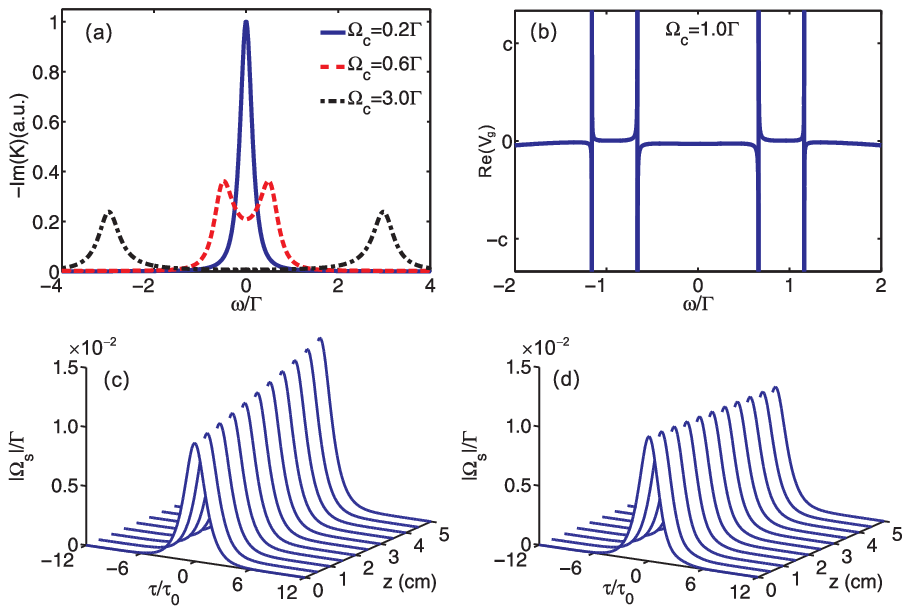}
\caption{(Color online) (a) Gain spectrum of the signal field,
$-$Im($K$), as a function of $\omega/\Gamma$ and $\Omega_c$ for the
GAN system. When $\Omega_c=0.2\Gamma$ it has only a single peak
(blue solid line). For $\Omega_c=0.6\Gamma$ a gain doublet appears
(red dashed line) and the doublet becomes wide by increasing
$\Omega_c$ ($=3.0\Gamma$) (black dotted-dashed line). (b) Group
velocity of the signal field, ${\rm Re}(V_g)$,  as a function of
$\omega/\Gamma$ for $\Omega_c=1.0\Gamma$, which can be smaller than
$c$ (subluminal), larger than $c$ and even negative (superluminal).
(c) ((d)\,) Numerical result for $|\Omega_s|/\Gamma$ as function of
$\tau/\tau_0$  ($\tau_0=1.0\times10^{-6}~\text{s}$) and $z$ for
$\Omega_c=0.2\Gamma$ ($\Omega_c=3.0\Gamma$).}\label{fig2}
\end{figure}

Shown in Fig.~\ref{fig2}(a)
%
%
is the gain spectrum of the signal field, i.e. $-$Im($K$), as a function of $\omega/\Gamma$ and
$\Omega_c$. red When plotting the figure, we have chosen the atomic vapor of $^{85}$Rb with the
states assigned by $|1\rangle=|5^2S_{1/2},F=2,m_F=-1\rangle$,
$|2\rangle=|5^2S_{1/2},F=3,m_F=1\rangle$, $|3\rangle=|5^2P_{1/2},F=2,m_F=0\rangle$, and
$|4\rangle=|5^2P_{3/2},F=4,m_F=2\rangle$. The system parameters are given by
$\Gamma=6~\text{MHz}$, $\Gamma_3=\Gamma_4=\Gamma$,
$\Gamma_{13}=\Gamma_{23}=\Gamma_{14}=\Gamma_{24}=\Gamma/2$, $\gamma_{21}=10~\text{kHz}$,
$\Delta_2=\Delta_4=0$, $\Delta_3=0.5~\text{GHz}$, $\Omega_{p}=1.0 \Gamma$, and
$\kappa_{23}/\Gamma=10^3\,{\rm cm}^{-1}$.  From the figure we see that when the control field is weak ($\Omega_c=0.2\Gamma$)
the gain spectrum  has only a single peak near $\omega=0$  (blue solid line). Interestingly, the
single peak will becomes into two peaks if the condition
\begin{equation}\label{hc-N}
|\Omega_c|^2 > \frac{\kappa_{23}|\Omega_p|^2\gamma_{41}}{\Delta^2_3}
\end{equation}
is satisfied. Illustrated by the red dashed line in Fig.~\ref{fig2}(a) is for
$\Omega_{c}=0.6\Gamma$. In this case the condition (\ref{hc-N}) is fulfilled and hence
a gain doublet (a dip between two gain peaks) appears in the gain spectrum.
By increasing $\Omega_{c}$ to $3.0\Gamma$, the distance between the two
peaks becomes wide and the gain of the signal field is nearly
vanishing at $\omega=0$ (black dotted-dashed line). The appearance
of the gain doublet in the gain spectrum is due to a QI effect in
the system, which will be explained in the next subsection.

Fig.~\ref{fig2}(b) shows the group velocity ${\rm Re}(V_g)$ of the
signal pulse as a function of $\omega/\Gamma$ when control field
$\Omega_c=1.0\Gamma$. The system parameters used are the same as in
Fig.~\ref{fig2}(a). We see that ${\rm Re}(V_g)$ can be smaller than
$c$ (subluminal), larger than $c$ and even negative (superluminal).
Especially, at $\omega=0$ we have ${\rm
Re}(V_g)=-0.51\times10^{-3}c$. Hence the atomic system with GAN
level configuration is indeed a fast-light medium.

In addition to acquire QI and hence the gain doublet in the gain
spectrum, the introduction of the control field can also stabilize
the propagation of the signal pulse. Shown in Fig.~\ref{fig2}(c) is
the numerical result based on the MB Eqs.~(\ref{BEGAN}) and
(\ref{Maxwell-N}) for $|\Omega_s|/\Gamma$ as function of
$\tau/\tau_0$ and $z$ for $\Omega_c=0.2\Gamma$, where $\tau=t-z/{\rm
Re}(V_g)$ is traveling coordinate and
$\tau_0=1.0\times10^{-6}~\text{s}$ is initial pulse width. In this
case (weak control field), the signal pulse has a large gain and
hence a significant deformation happens during propagation. However,
for a large control field ($\Omega_c=3.0\Gamma$), the gain is
largely suppressed and hence no deformation occurs during
the propagation of the signal pulse. In the numerical simulation, we
have taken the initial signal pulse with the form
$\Omega_s=0.01\Gamma,\text{sech}(\tau/\tau_0)$. From these results
we conclude that the GAN system with a large control field can
support a stable propagation of the signal field.

\subsection{Crossover from constructive QI to ATS in the GAN system}\label{SecIIb}

Now we make a detailed analysis on the appearance of gain doublet in the
gain spectrum of the signal field shown above by using the
SDM developed recently for slow-light
media~\cite{Agarwal1997,Abi-Salloum2010,Anisimov2011,
Giner2013,Tan2013,Zhu2013,Tan2014,Peng2014,Ding,Lu,Davuluri}.

To this end, we simplify the linear dispersion relation (i.e. Eq.~(\ref{K-N})\,).
Under the condition $\Delta_3 \gg \gamma_{ij}, \Gamma_{ij}, \Omega_{p},
\Omega_{c}$, Eq.~(\ref{K-N}) for $\Delta_2=\Delta_4=0$ can be reduces
to the form
$K(\omega)=\omega/c+\widetilde{\kappa}_{23}(\omega
+i\gamma_{41})/[(\omega+i\gamma_{21})(\omega
+i\gamma_{41})-|\Omega_{c}|^2]$,
with
$\widetilde{\kappa}_{23}=\kappa_{23}\Omega_{p}\sigma_{31}^{(0)*}/\Delta_3$.
One can also obtain a similar expression of $K(\omega)$ for
nonvanishing $\Delta_2$ and $\Delta_4$, but it is lengthy and thus
omitted here. $K(\omega)$ can be written into the form
\begin{equation}\label{Kd1}
K(\omega)=\frac{\omega}{c}+\widetilde{\kappa}_{23}\frac{\omega
+i\gamma_{41}}{(\omega-\omega_+)(\omega-\omega_-)},
\end{equation}
where $\omega_{\pm}=-i(\gamma_{21}+\gamma_{41})/2 \pm \left[\left(
|\Omega_{c}|^2-|\Omega_{\rm ref}|^2\right)\right]^{1/2}$, with
$\Omega_{\rm ref}=|\gamma_{21}-\gamma_{41}|/2$  (reference half
Rabi frequency).

In order to illustrate the QI effect in a simple and clear way, as done in
Refs.~\cite{Agarwal1997,Abi-Salloum2010,Anisimov2011,
Giner2013,Tan2013,Zhu2013,Tan2014,Peng2014,Ding,Lu,Davuluri},
we decompose the gain spectrum $-\text{Im}(K)$ for different regions of $\Omega_{c}$
as follows.

(i). {\it Weak control field region} ($|\Omega_{c}|<\Omega_{\rm
ref}\approx\gamma_{14}/2$): Equation (\ref{Kd1}) can be decomposed
as
\begin{equation}\label{}
K(\omega)=\frac{\omega}{c}+\widetilde{\kappa}_{23}\left(\frac{A_+}{\omega
-\omega_+}+\frac{A_-}{\omega-\omega_-}\right),
\end{equation}
with $A_{\pm}=\pm(\omega_{\pm}+i\gamma_{41})/(\omega_+-\omega_-)$.
Since in this region ${\rm Re}(\omega_{\pm})={\rm Im}(A_{\pm})=0$,
we obtain the gain spectrum
\begin{equation}\label{form1}
-{\rm
Im}(K)=-\widetilde{\kappa}_{23}\left(\frac{B_+}{\omega^2+\delta_+^2}+
\frac{B_-}{\omega^2+\delta_-^2}\right)= -L_+ - L_-,
\end{equation}
where $\delta_{\pm}={\rm Im}(\omega_{\pm})$,
$B_{\pm}=A_{\pm}\delta_{\pm}$, and
$L_{\pm}=\widetilde{\kappa}_{23}B_{\pm}/(\omega^2+\delta_{\pm}^2)$.
Thus the signal field gain spectrum comprises two Lorentzians
centered at $\omega=0$.

\begin{figure}
\includegraphics[scale=0.63]{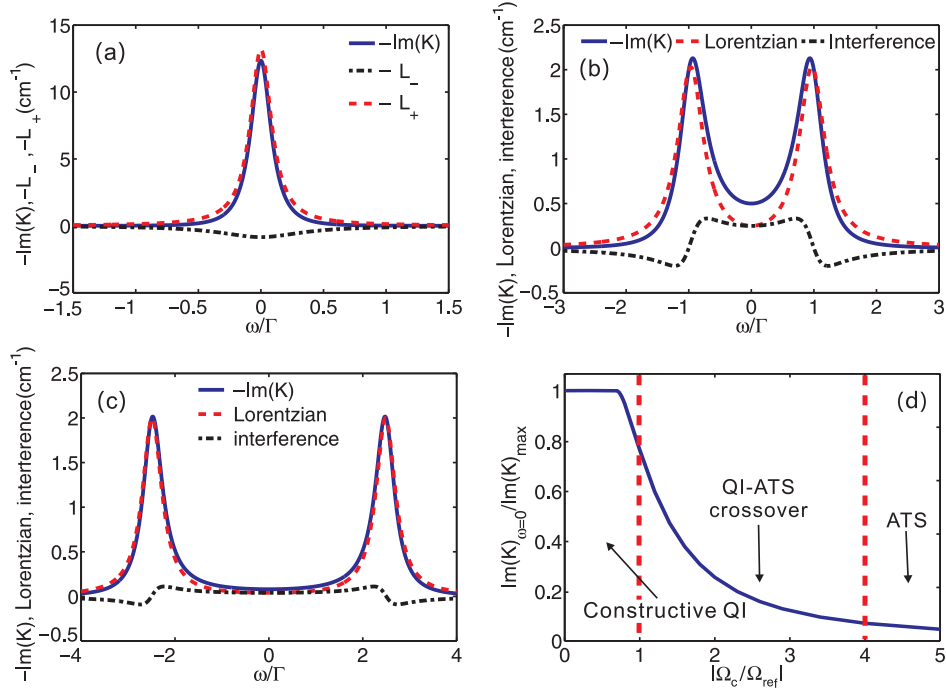}
\caption{(Color online) (a) $-$Im($K$) in the weak control field
region ($\Omega_{c}=0.2\Gamma<\Omega_{\rm ref}$) as a function of
$\omega/\Gamma$. The red dashed (the black
dashed-dotted) line is for $-L_+$ ($-L_-$); the blue solid line is for
$-$Im($K$). (b) Gain spectrum in the intermediate control field
region ($\Omega_{c}=1.0\Gamma<\Omega_{\rm ref}$). Red dashed line is
for two Lorentzian terms; black dashed-dotted line is for
constructive QI term; blue solid line is for $-$Im($K$).　 (c) Gain
spectrum in the strong control field region
($\Omega_{c}=2.5\Gamma>\Omega_{\rm ref}$). Red dashed line is for
two Lorentzian terms; black dashed-dotted line is for small
constructive QI term; blue solid line is for $-$Im(K). (d) The
``phase diagram'' of Im$(K)_{\omega=0}/\text{Im}(K)_{\text{max}}$ as
a function of $|\Omega_{c}/\Omega_{\text{ref}}|$ illustrating the
transition from constructive QI to ATS in the GAN system. Three
regions (i.e. constructive QI region, the QI-ATS crossover region,
and the ATS region) are separated by two vertical dashed
lines.}\label{fig3}
\end{figure}

Shown in Fig.~\ref{fig3}(a) is the result of $-L_+$, which has a
large positive single peak (red dashed line), and $-L_-$, which has
a negative (concave but very shallow) single peak (black
dashed-dotted line). When plotting the figure, a small control field
($\Omega_{c}=0.2\Gamma$) is taken. We see that although for small
$\Omega_{c}$ there is a destructive interference between $-L_+$ and
$-L_-$, the interference is too small and hence can be neglected. So
the superposition of $-L_+$ and $-L_-$ is contributed mainly by
$-L_+$. As a result, the gain spectrum $-$Im($K$) displays only a
single peak (blue solid line).

%
%

(ii). {\it Intermediate control field region} ($|\Omega_{c}|>\Omega_{\rm
ref}$): In this region ${\rm Re}(\omega_{\pm})\neq0$, we obtain
\begin{eqnarray}\label{form2}\nonumber
-{\rm
Im}(K)=&& \frac{\widetilde{\kappa}_{23}}{2}\left\{\frac{W}{(\omega-\delta)^2+W^2}
+\frac{W}{(\omega+\delta)^2+W^2}\right.\nonumber\\
&&\left.\hspace{0.7cm}+\frac{g}{\delta}\left[\frac{\omega-\delta}{(\omega
-\delta)^2+W^2}-\frac{\omega+\delta}{(\omega+\delta)^2+W^2}\right]\right\},
\end{eqnarray}
with $W=(\gamma_{21}+\gamma_{41})/2$,
$\delta=\sqrt{4|\Omega_{c}|^2-(\gamma_{21}-\gamma_{41})^2}/2$,
and $g=(\gamma_{21}-\gamma_{41})/2$. The previous two terms in the
bracket $\{\cdots\}$ of the above expression (i.e. the two
Lorentzian terms) can be thought as the net contribution
contributed to the gain resonance from two different channels
corresponding to the two dressed states (i.e. the states $|2\rangle$
and $|4\rangle$) created by the control field $\Omega_{c}$. The term
proportional to $g$ is clearly a QI term. The QI is controlled by
the parameter $g$ and it is destructive (constructive) if $g>0$
($g<0$). Since in the GAN system $\gamma_{21}$ is always much
smaller than $\gamma_{41}$ and $g$ is often negative, thus the QI in
the system is a constructive one.

Fig.~\ref{fig3}(b) shows the gain spectrum in the intermediate
control field region ($|\Omega_{c}|=1.0 \Gamma >\Omega_{\rm ref}$).
The red dashed line is for the two Lorentzian terms; the black
dashed-dotted line is for the constructive QI term; the blue solid
line is for $-$Im($K$).　 We see that in this region a gain doublet
appears, which is the result of the superposition of the two
Lorentzian terms and the QI term.

(iii). {\it Large control field region} ($|\Omega_{c}|\gg\Omega_{\rm
ref}$): In this case, the gain spectrum is still given by
(\ref{form2}), but the strength of the QI term, $g/\delta$, is very
weak (i.e. $g/\delta \approx 0$). Thus one has
\begin{equation}\label{form3}
-{\rm
Im}(K)\approx \frac{\widetilde{\kappa}_{23}}{2}\left[\frac{W}{(\omega-\delta)^2
+W^2}+\frac{W}{(\omega+\delta)^2+W^2}\right].
\end{equation}

Shown in Fig.~\ref{fig3}(c) is the gain
spectrum as a function of $\omega/\Gamma$ for
$|\Omega_{c}|=2.5\Gamma \gg \Omega_{\rm ref}$. The red dashed line
represents the contribution by the sum of the two Lorentzian terms.
For illustration, we have also plotted the contribution from the
small QI term (omitted in Eq.~(\ref{form3})\,), denoted by the black
dotted-dashed line. We see that in this case the QI is small and
weak constructive. The blue solid line is the curve of
$-\text{Im}(K)$, which has two resonance peaks at
$\omega\approx\pm\Omega_{c}$. Obviously, the phenomenon found in
this case belongs to ATS because the window of the gain doublet is
mainly due to the contribution by the two Lorentzian
terms. From the results given in Fig.~\ref{fig3}(b) and
Fig.~\ref{fig3}(c), we conclude that the GAN system possesses indeed
QI effect and allows a crossover from the QI to the
ATS.

Fig.~\ref{fig3}(d) shows the ``phase diagram'' that illustrates the
crossover from the constructive QI effect to the ATS in the GAN
system by plotting
$\text{Im}(K)_{\omega=0}/\text{Im}(K)_{\text{max}}$ as a function of
$|\Omega_{c}/\Omega_{\text{ref}}|$. We see that the phase diagram
can be roughly divided into three regions, i.e. the constructive QI
region (weak control field region), the QI-ATS crossover region
(intermediate control field region), and the ATS region (large
control field region), very similar to that found in various EIT
systems reported recently for slow
lights~\cite{Agarwal1997,Abi-Salloum2010,Anisimov2011,
Giner2013,Tan2013,Zhu2013,Tan2014,Peng2014,Ding,Lu,Davuluri}.

The physical reason for the occurrence of the QI (i.e. the gain doublet in the gain spectrum
in Fig.~\ref{fig3}(a)\,) can be  explained as follows. From the Hamiltonian give above
we obtain an eigenstate of the system
\begin{equation}\label{DARK_N}
|\psi\rangle=\left[1+\frac{|\Omega_s|^2d_4}{(|\Omega_c|^2-d_2d_4)d_3}\right]|1\rangle
-\frac{\Omega_p\Omega^*_sd_4}{(|\Omega_c|^2-d_2d_4)d_3}|2\rangle-
\frac{\Omega_p}{d_3}|3\rangle+
\frac{\Omega_p\Omega_c\Omega^*_s}{(|\Omega_c|^2-d_2d_4)d_3}|4\rangle,
\end{equation}
with $d_2=\Delta_2+i\Gamma_2$, $d_3=\Delta_3+i\Gamma_3$, $d_4=\Delta_4+i\Gamma_4$.
For large $\Delta_3$ and $\Omega_c$ and for small $\Delta_2$ and $\Delta_4$,
the eigenstate (\ref{DARK_N})  becomes
$|\psi \rangle \approx |1\rangle-(\Omega_p/\Delta_3)|3\rangle+[\Omega_p\Omega_c\Omega^*_s/(|\Omega_c|^2 \Delta_3)]\,|4\rangle$,
i.e. the atomic state $|2 \rangle $ is not involved. Such state is a ``dark state'' with zero eigenvalue resulted from
the quantum interference  between the transition paths $|3\rangle\rightarrow |2\rangle$ and $|4\rangle\rightarrow |2\rangle$.
As a result, the gain is largely suppressed
and hence the gain doublet appears in the gain spectrum of the signal field.

\section{Quantum interference characters of the GAL-I system}\label{Sec:3}

\subsection{Model and linear dispersion relation}

We now turn to consider the GAL-I system shown in
Fig.~\ref{fig1}(b).
The differences between the GAL-I system and the GAN system (Fig.~\ref{fig1}(a)\,)
are that in the GAL-I system the state $|4\rangle$ is below the state
$|2\rangle$. It is just these two differences that makes the QI character
of the GAL-I system very different from that in the GAN system, as
shown below.

Under EDA and RWA, in interaction picture the Hamiltonian of the GAL-I system is given by
$\hat{\cal H}=-\hbar[\sum_{j=1}^4 \Delta_j|j\rangle\langle
j|+(\Omega_p|3\rangle\langle1|+\Omega_s|3\rangle\langle2|+\Omega_{c}|2\rangle\langle4|+{\rm
h.c.})] $, where the definitions of $\Omega_p$  and $\Omega_s$ are
the same as those in the GAN system, but now $\Omega_{c}
=(\boldsymbol{p}_{24}\cdot{\cal E}_{c})/\hbar$. The Bloch equation
of the system is presented in Appendix~\ref{appA2}. The equation of
motion for $\Omega_s$ can be derived by the Maxwell equation under
SVEA, which reads
\begin{equation}\label{Maxwell-N2}
i\left(\frac{\partial}{\partial
z}+\frac{1}{c}\frac{\partial}{\partial t}\right)\Omega_s
+\kappa_{23}\sigma_{32}=0,
\end{equation}
with $\kappa_{23}={\cal
N}_a\omega_s|\boldsymbol{p}_{32}|^2/(2\hbar\varepsilon_0 c)$.

The base state solution of the system can be obtained by using the MB
equations Eqs.~({\ref{Ladder-Ia}), Eqs.~(\ref{Ladder-Ib}) and Eq.~(\ref{Maxwell-N2}).
For large $\Delta_3$, it reads
$\sigma_{11}^{(0)}\approx 1$, $\sigma_{31}^{(0)}\approx -\Omega_p/d_{31}$,
and all other $\sigma_{jl}^{(0)}\approx 0$ (see Appendix~\ref{appA2}), i.e.
the atomic medium is initially prepared with the population mainly
in the ground state state $|1\rangle$.
When the weak signal field is applied, the solution at the first order of $\Omega_s$
reads
\begin{subequations}\label{First-I}
\begin{eqnarray}
&&\Omega_s=Fe^{i\theta},\\
&&\sigma_{32}^{(1)}=\frac{B(\sigma_{33}^{(0)}-\sigma_{22}^{(0)})-
(D_{p}+|\Omega_{c}|^2)\Omega_{p}\sigma_{31}^{*(0)}-(D_{c}
+|\Omega_{p}|^2)\Omega_{c}\sigma_{42}^{*(0)}}{(\omega+d_{32})B-
|\Omega_{p}|^2(D_{p}+|\Omega_{c}|^2)-|\Omega_{c}|^2(D_{c}+|\Omega_{p}|^2)}Fe^{i\theta},
\end{eqnarray}
\end{subequations}
where $D_{p}=(\omega-d_{41}^*)(\omega-d_{43}^*)-|\Omega_{p}|^2$,
$D_{c}=(\omega-d_{21}^*)(\omega-d_{41}^*)-|\Omega_{c}|^2$ and
$B=(\omega-d_{21}^*)(\omega-d_{41}^*)(\omega-d_{43}^*)-(\omega-d_{21}^*)|\Omega_{p}|^2
-(\omega-d_{43}^*)|\Omega_{c}|^2$.  The linear dispersion relation of the system reads
\begin{equation}\label{K-I}
K=\frac{\omega}{c}+\kappa_{23}\frac{B(\sigma_{33}^{(0)}-\sigma_{22}^{(0)})
-(D_{p}+|\Omega_{c}|^2)\Omega_{p}\sigma_{31}^{*(0)}-(D_{c}
+|\Omega_{p}|^2)\Omega_{c}\sigma_{42}^{*(0)}}{(\omega+d_{32})B-|\Omega_{p}|^2(D_{p}+|\Omega_{c}|^2)
-|\Omega_{c}|^2(D_{c}+|\Omega_{p}|^2)}.
\end{equation}

In Fig.~\ref{fig4}(a)
%
\begin{figure}
\includegraphics[scale=0.65]{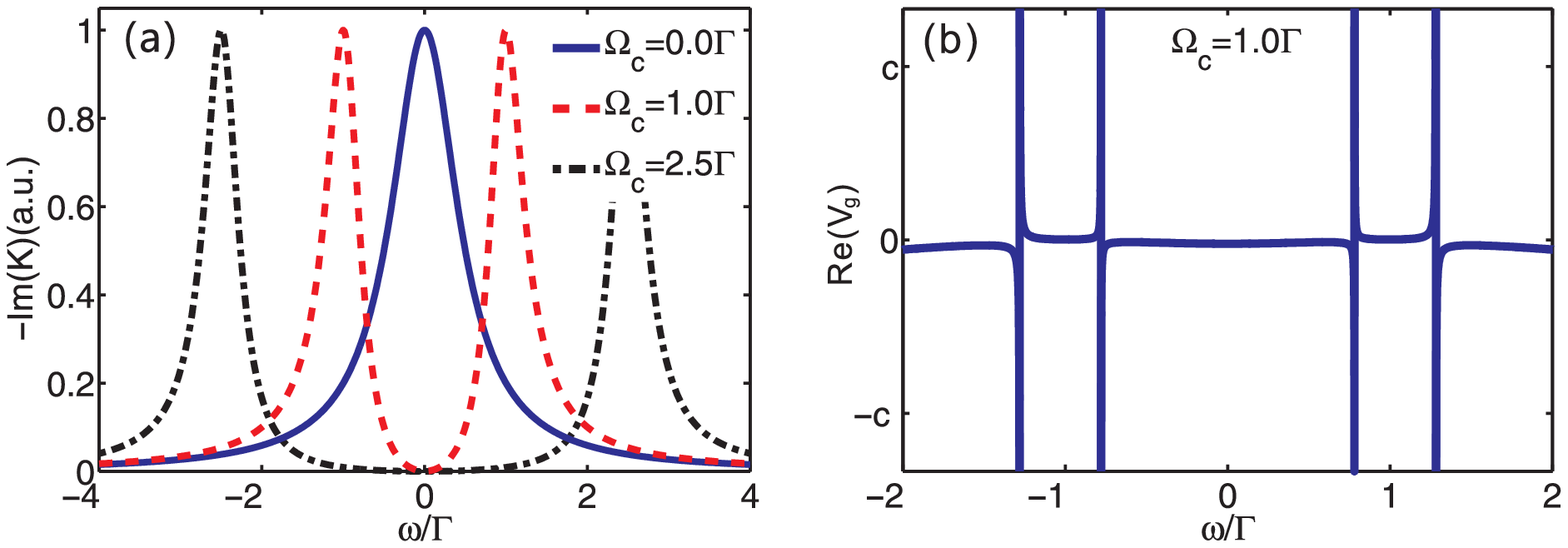}
\caption{(Color online)
(a) Gain spectrum of the signal field, $-$Im($K$), as a
function of $\omega/\Gamma$ and $\Omega_c$ for the GAL-I system.
For $\Omega_c=0$, it has only a single peak
(blue solid line). For $\Omega_c=1.0\Gamma$, a gain doublet appears
(red dashed line) and the doublet becomes wide when
$\Omega_c=2.5\Gamma$ (black dotted-dashed line).
(b) Group velocity of the signal field, ${\rm Re}(V_g)$,  as a function of
$\omega/\Gamma$  for $\Omega_c=1.0\Gamma$, which can be smaller than $c$ (subluminal),
larger than $c$ and even negative (superluminal).
}\label{fig4}
\end{figure}
%
we show the gain spectrum of the GAL-I system, i.e. $-$Im($K$), as a function of $\omega/\Gamma$
and $\Omega_c$. States $|1\rangle$ and $|3\rangle$ in the system are assumed to be coupled with a
pump field through a two-photon transition (with effective half Rabi frequency $\Omega_p$). When
plotting the figure we also use the atomic vapor of $^{85}$Rb, with the levels assigned as
$|1\rangle=|5^2S_{1/2},~F=2,~m_F=-1\rangle$, $|2\rangle=|5^2P_{1/2},~F=2,~m_F=0\rangle$,
$|3\rangle=|5^2D_{5/2},~F=3,~m_F=-1\rangle$, and $|4\rangle=|5^2S_{1/2},~F=3,~m_F=1\rangle$. The
system parameters used are $\Gamma=6~\text{MHz}$, $\Gamma_3=\Gamma$,
$\Gamma_{13}=\Gamma_{23}=\Gamma_{42}=0.5\Gamma$, $\gamma_{41}=10~\text{kHz}$,
$\Delta_2=\Delta_4=0$, $\Delta_3=0.5~\text{GHz}$, $\Omega_{p}=0.5 \Gamma$, and
$\kappa_{23}/\Gamma=10^3$ cm$^{-1}$.
One sees that for $\Omega_c=0$ the gain
spectrum has only a single peak (blue solid line). Increasing $\Omega_c$ to satisfy the condition
$|\Omega_c|^2
> \kappa_{23}|\Omega_p|^2 \gamma_{41}/\Delta^2_3$ (the same as
the condition (\ref{hc-N})\,), a gain doublet appears. The red dashed line in the
figure is the result for $\Omega_c=1.0\Gamma$. When
$\Omega_c=2.5\Gamma$, the gain doublet becomes wide. The occurrence
of the gain doublet is due to a QI effect in the system.

By inspection of Fig.~\ref{fig2}(a) and Fig.~\ref{fig4}(a), we find that there are some similar
physical characters between them. First, when $\Omega_c$ is small, Gain spectrum has only a single
peak. Second, increasing $\Omega_c$ to a critical value (i.e. $\Omega_c > [\kappa_{23}|\Omega_p|^2
\gamma_{41}/\Delta^2_3]^{1/2}$) the single peak becomes two peaks. By increasing $\Omega_c$
further, the width between the two peaks increases. However, there are some differences between
Fig.~\ref{fig2}(a) and Fig.~\ref{fig4}(a). The most obvious one is that the peak values of
$-$Im($K$) in Fig.~\ref{fig2}(a) are reduced rapidly when $\Omega_c$ is increased, but such
phenomenon does not occur in Fig.~\ref{fig4}(a). One of physical reasons is that in the GAN system
(Fig.~\ref{fig2}(a)\,) the level $|4\rangle$ is above the level $|2\rangle$ and the large
$\Gamma_{24}$ (the spontaneous emission rate from $|4\rangle$ to $|2\rangle$) contributes a
population to $|2\rangle$ and hence reduces the gain of the signal field significantly.
Differently, in the GAL-I system (Fig.~\ref{fig4}(a)\,) the level $|4\rangle$ is below the level
$|2\rangle$ and $\Gamma_{42}$ (the spontaneous emission rate from $|2\rangle$ to $|4\rangle$) is
small and hence it has no significant influence to the gain spectrum. As a result, the peak values
in the gain spectrum of the GAL-I system has no significant change as $\Omega_c$ is increased.

Fig.~\ref{fig4}(b) shows the group velocity of the signal pulse as a
function of $\omega/\Gamma$ for $\Omega_c=1.0\Gamma$. The system
parameters used for plotting the figure are the same as those in
Fig.~\ref{fig4}(a). One sees that ${\rm Re}(V_g)$ can be smaller and
larger than $c$, and even negative, and hence the GAL-I
system is a typical fast-light medium.

\subsection{Crossover from destructive QI to ATS in the GAL-I system}\label{SecIIIb}

We now employ the SDM used in Sec.~\ref{SecIIb}
to make a detailed analysis on the QI effect in the GAL-I system.
For $\Delta_3$ much larger than $\gamma_{ij}$, $\Gamma_{ij}$, $\Omega_{p}$,
and $\Omega_{c}$, Eq.~(\ref{K-I}) is reduced to
$K=\omega/c+\widetilde{\kappa}_{23}(\omega+i\gamma_{41})/[(\omega+i\gamma_{21})
(\omega+i\gamma_{41})-|\Omega_{c}|^2]$
with $\widetilde{\kappa}_{23}=\kappa_{23}\Omega_{p}\sigma_{31}^{(0)*}/\Delta_3$,
which can be written as the form
\begin{equation}\label{K-dI}
K(\omega)=\frac{\omega}{c}+\widetilde{\kappa}_{23}\frac{\omega
+i\gamma_{41}}{(\omega-\omega_+)(\omega-\omega_-)}.
\end{equation}
Here
$\omega_{\pm}=-i(\gamma_{21}+\gamma_{41})/2 \pm \left[
|\Omega_{c}|^2-|\Omega_{\rm ref}|^2\right]^{1/2}$
with $\Omega_{\rm ref}=|\gamma_{21}-\gamma_{41}|/2$ the reference half Rabi frequency.
The gain spectrum $-\text{Im}(K)$ can be decomposed into three different regions.

(i). {\it Weak control field region} ($|\Omega_{c}|<\Omega_{\rm
ref}$):  Eq.~(\ref{K-dI}) can be decomposed into
\begin{equation}\label{Kd2}
K(\omega)=\frac{\omega}{c}+\widetilde{\kappa}_{23}\left(\frac{A_+}{\omega
-\omega_+}+\frac{A_-}{\omega-\omega_-}\right),
\end{equation}
where $A_{\pm}=\pm(\omega_{\pm}+i\gamma_{41})/(\omega_+-\omega_-)$.
Since ${\rm Re}(\omega_{\pm})={\rm Im}(A_{\pm})=0$, we have
\begin{equation}\label{form1'}
-{\rm
Im}(K)=-\widetilde{\kappa}_{23}\left(\frac{B_+}{\omega^2+\delta_+^2}+
\frac{B_-}{\omega^2+\delta_-^2}\right)=-L_+-L_-,
\end{equation}
with $\delta_{\pm}={\rm Im}(\omega_{\pm})$,
$B_{\pm}=A_{\pm}\delta_{\pm}$, and
$L_{\pm}=\widetilde{\kappa}_{23}B_{\pm}/(\omega^2+\delta_{\pm}^2)$.
Thus the signal-field gain profile comprises two
Lorentzian centered at the origin ($\omega=0$).

Fig.~\ref{fig5}(a) shows
%
\begin{figure}
\includegraphics[scale=1.05]{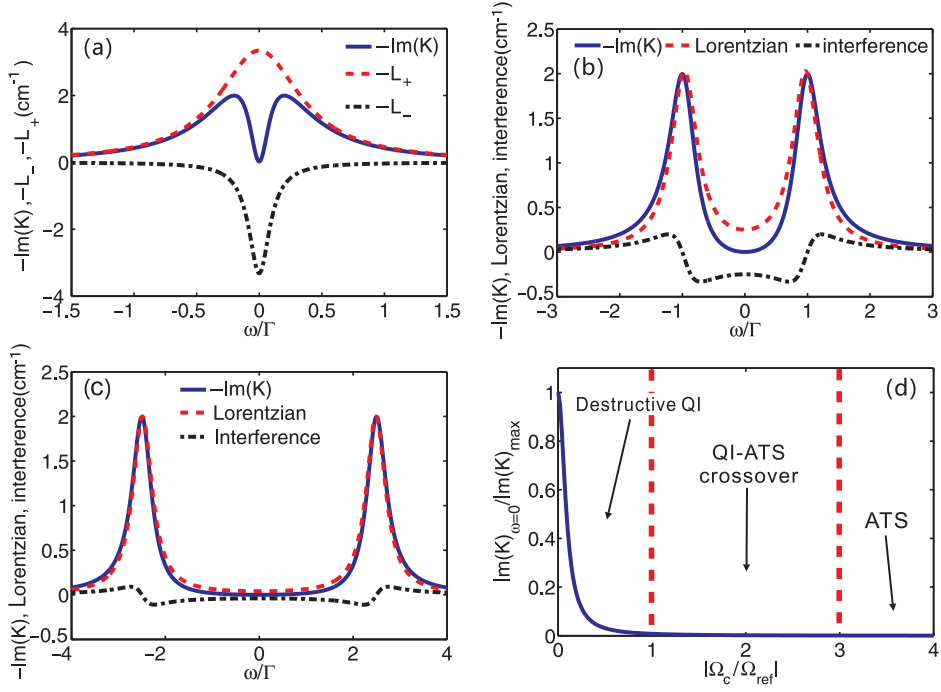}
\caption{(Color online) (a) Gain spectrum  $-$Im($K$) in the weak control field
region ($\Omega_{c}=0.2\Gamma<\Omega_{\rm ref}$) as a function of
$\omega/\Gamma$. Red dashed line: $-L_+$; black dashed-dotted line:
$-L_-$; blue solid line: $-$Im($K$). (b) Gain spectrum in the
intermediate control field region ($\Omega_{c}=1.0\Gamma<\Omega_{\rm
ref}$). Red dashed line: two Lorentzian terms; black dashed-dotted
line: destructive QI term; blue solid line: $-$Im($K$).　 (c) Gain
spectrum in the strong control field region
($\Omega_{c}=2.5\Gamma>\Omega_{\rm ref}$). Red dashed line: two
Lorentzian terms; black dashed-dotted line: destructive QI term;
blue solid line: $-$Im(K). (d) The ``phase diagram'' of
Im$(K)_{\omega=0}/\text{Im}(K)_{\text{max}}$ as a function of
$|\Omega_{c}/\Omega_{\text{ref}}|$ illustrating the crossover from
destructive QI  to ATS in the GAL-I system. Three regions (i.e.
destructive QI region, QI-ATS crossover region, and ATS region) are
divided by two vertical dashed lines.}\label{fig5}
\end{figure}
%
the profile of $-L_+$, which is a
positive single peak (red dashed line), and $-L_-$, which is a
negative single peak (black dash-dotted line). When plotting the
figure $\Omega_{c}=0.2\Gamma$ is chosen and the other system parameters are
the same as in Fig.~\ref{fig4}(a). The superposition of
$-L_+$ and $-L_-$ gives the profile of $-\text{Im}(K)$ (blue solid line),
which has a gain doublet with a gain window opened near $\omega=0$.
We see that there exists a destructive QI in the GAL-I system,
reflecting by the sum of the positive $-L_+$ and the negative
$-L_-$, very similar to the destructive QI in the ladder-I type EIT system
found recently~\cite{Tan2014}.

(ii). {\it  Intermediate control field region} ($|\Omega_{c}|>\Omega_{\rm
ref}$): In this region ${\rm Re}(\omega_{\pm})\neq0$, we obtain
\begin{eqnarray}\label{form2-I}\nonumber
-{\rm
Im}(K)=&&\frac{\widetilde{\kappa}_{23}}{2}\left\{\frac{W}{(\omega-\delta)^2+W^2}
+\frac{W}{(\omega+\delta)^2+W^2}\right.\nonumber\\
&&\left.\hspace{0.7cm}+\frac{g}{\delta}\left[\frac{\omega-\delta}{(\omega
-\delta)^2+W^2}-\frac{\omega+\delta}{(\omega+\delta)^2+W^2}\right]\right\},
\end{eqnarray}
with $W=(\gamma_{21}+\gamma_{41})/2$,
$\delta=[4|\Omega_{c}|^2-(\gamma_{21}-\gamma_{41})^2]^{1/2}/2$,
and $g=(\gamma_{21}-\gamma_{41})/2$. The first two terms
(Lorentzian terms) in bracket of Eq.~(\ref{form2-I})
are the net contribution by the gain
resonance from two different channels corresponding to the two
dressed states (state $|2\rangle$ and $|4\rangle$). The
terms proportional to $g$ are clearly interference ones. The
interference is governed by the parameter $g$ and it is
destructive (constructive) if $g>0$ ($g<0$). Because in the GAL-I
system $\gamma_{21}$ is much smaller than $\gamma_{41}$, $g$ is
always positive. Thus the QI induced by the control field is
destructive.

Fig.~\ref{fig5}(b) shows the gain spectrum $-$Im($K$) in the intermediate control field region.
The red dashed line is for the two positive Lorentzian terms; the black dashed-dotted line is for
the negative QI term. We see that in this region a gain doublet appears in the gain spectrum (blue
solid line), which is the result of the superposition of the two positive Lorentzian terms and the
negative (thus destructive) QI term. When plotting the figure, the system parameters used are the
same as those in Fig.~\ref{fig5}(a) except $\Omega_{c}=1.0 \Gamma$.

(iii). {\it Large control field region} ($|\Omega_{c}|\gg\Omega_{\rm ref}$): In this case, the
quantum interference strength $g/\delta$ in Eq.~(\ref{form2-I}) is very weak (i.e.
$g/\delta\approx 0$), and hence Im$(K)$ can be simplified to $-{\rm
Im}(K)=(\widetilde{\kappa}_{23}/2)\left\{W/[(\omega-\delta)^2
+W^2]+W/[(\omega+\delta)^2+W^2]\right\}$.
Shown in Fig.~\ref{fig5}(c) is the signal-field gain spectrum as a function of $\omega/\Gamma$ in
the large control field region ($\Omega_{c}=2.5\Gamma$). The red dashed line is the contribution
by the sum of the two positive Lorentzian terms. The contribution of very small interference terms
are also plotted as the black dotted-dashed line, which is negative and thus destructive. The blue
solid line is for $-$Im($K$). Obviously, the phenomenon found in this case belongs to ATS because
the gain window is wide and mainly due to the contribution of two Lorentzian terms.

In Fig.~\ref{fig5}(d) we show the ``phase diagram'' of the system, which reflects the crossover
from the destructive QI effect to the ATS in the GAL-I system, by taking
$\text{Im}(K)_{\omega=0}/\text{Im}(K)_{\text{max}}$ as a function of
$|\Omega_{c}/\Omega_{\text{ref}}|$. We see that the phase diagram can also be divided into three
regions, i.e. the destructive QI region (weak control field region), the QI-ATS crossover region
(intermediate control field region), and the ATS region (large control field region), similar to
those found in EIT systems for slow lights~\cite{Agarwal1997,Abi-Salloum2010,Anisimov2011,
Giner2013,Tan2013,Zhu2013,Tan2014,Peng2014,Ding,Lu,Davuluri}.

The physical reason for the occurrence of the QI here (i.e. the gain doublet in the gain spectrum
in Fig.~\ref{fig4}(a)\,) can also be explained by the existence of a dark state in the system.
From the system Hamiltonian we obtain an eigenstate
\begin{equation}\label{DARK_L}
|\psi\rangle=\left[1+\frac{|\Omega_s|^2d_4}{(|\Omega_c|^2-d_2d_4)d_3}\right]|1\rangle-
\frac{\Omega_p\Omega_sd_4}{(|\Omega_c|^2-d_2d_4)d_3}|2\rangle-
\frac{\Omega_p}{d_3}|3\rangle+
\frac{\Omega_p\Omega^*_c\Omega_s}{(|\Omega_c|^2-d_2d_4)d_3}|4\rangle,
\end{equation}
where $d_2=\Delta_2+i\Gamma_2$, $d_3=\Delta_3+i\Gamma_3$, $d_4=\Delta_4+i\Gamma_4$. For large
$\Delta_3$ and $\Omega_c$ and for small $\Delta_2$ and $\Delta_4$, the eigenstate (\ref{DARK_L})
reduces to $|\psi \rangle \approx
|1\rangle-(\Omega_p/\Delta_3)|3\rangle+[\Omega_p\Omega_c^*\Omega_s/(|\Omega_c|^2
\Delta_3)]\,|4\rangle$. Because state $|2 \rangle $ is not involved, $|\psi \rangle$ is a ``dark
state'' with zero eigenvalue, which is resulted from the interference between the two quantum
transition passages from $|3\rangle\rightarrow |2\rangle$ and $|4\rangle\rightarrow |2\rangle$.
The outcome of such quantum interference is appearance of the gain doublet in the gain spectrum of
the signal field.

\section{Autler-Townes splitting in the GAL-II system}\label{Sec:4}

Finally, we consider the GAL-II system (Fig.~\ref{fig1}(c)). Under EDA and RWA, the Hamiltonian of
the GAL-II system in interaction picture is} $\hat{\cal H}=-\hbar[\sum_{j=1}^4
\Delta_j|j\rangle\langle j|+(\Omega_p|3\rangle\langle1|
+\Omega_c|3\rangle\langle2|+\Omega_s|2\rangle\langle4|+{\rm h.c.})]$. The Bloch equations for the
GAL-II system can be obtained from those of the GAL-I system given in Appendix~{\ref{appA2}, by
using the exchange of $\Omega_s\rightleftharpoons\Omega_c$. Under SVEA, the Maxwell equation of
the signal field reduces to
$ i\left[ \partial/\partial z+(1/c)\partial/\partial t\right]\Omega_s
+\kappa_{24}\sigma_{42}=0$,
with $\kappa_{24}={\cal
N}_a\omega_s|\boldsymbol{p}_{42}|^2/(2\hbar\varepsilon_0 c)$.  The base state of the system
is presented in Appendix~\ref{appA3}.

It is easy to obtain the linear dispersion relation of the system
\begin{equation}\label{K-II}
K(\omega)=\frac{\omega}{c}+\kappa_{24}\frac{(\omega+d_{41})\Omega_{\text{c}}\sigma^{(0)*}_{32}+
\Omega^*_{p}\Omega_{\text{c}}\sigma^{(0)}_{21}+D_{p}(\sigma_{44}^{(0)}
-\sigma_{22}^{(0)})}{D_{p}(\omega+d_{42})+|\Omega_{\text{c}}|^2(\omega+d_{41})},
\end{equation}
with $D_{p}=|\Omega_{p}|^2-(\omega+d_{41})(\omega+d_{43})$ and
$D_{c}=|\Omega_{c}|^2-(\omega+d_{42})(\omega+d_{43})$.

Shown in Fig.~\ref{fig6}(a)
%
\begin{figure}
\includegraphics[scale=0.66]{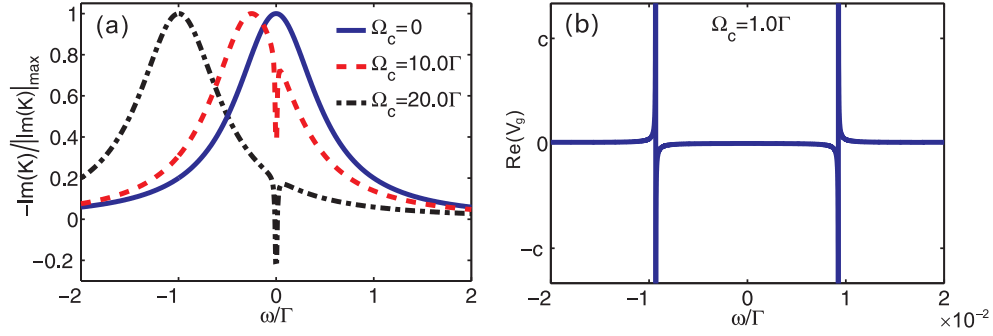}
\caption{(Color online) (a) Gain spectrum of the signal field,
$-$Im($K$), as a function of $\omega/\Gamma$ and $\Omega_c$ for the
GAL-II system. For $\Omega_c=0$ it has only a large single peak
centered at $\omega=0$ (blue solid line). For $\Omega_c=10.0\Gamma$, a very narrow gain doublet (dip)
is opened around $\omega=0$  and the large single peak moves to the left (red dashed line).
For $\Omega_c=20.0\Gamma$, the gain doublet still keeps to exist near at $\omega=0$ but the
the large single peak moves further to the left (black dotted-dashed line).
(b) Group velocity ${\rm Re}(V_g)$ of the signal field of the GAL-II system as a function of
$\omega/\Gamma$ for $\Omega_c=1.0\Gamma$.}\label{fig6}
\end{figure}
%
is the signal-field gain spectrum, $-$Im($K$), as a function of $\omega/\Gamma$ for different
$\Omega_c$. One sees that for $\Omega_c=0$ the gain spectrum has only a single peak centered at
$\omega=0$ (blue solid line); Increasing the value of $\Omega_c$ to $10.0\Gamma$, a gain doublet
(dip) opens around $\omega=0$ with the single peak moves to the left (red dashed line); Increasing
$\Omega_c$ further to $20.0\Gamma$, the gain doublet still keeps near at $\omega=0$ but the single
peak moves further to the left (black dotted-dashed line). When drawing the figure, the four
atomic states and the system parameters used are the same as those for the GAL-I system (see the
last section). The condition for the appearance of the gain doublet in the GAL-II system is
$|\Omega_c|^2>\Delta_3\gamma_{42}$.

Comparing Fig.~\ref{fig6}(a) with Fig.~\ref{fig2}(a) and Fig.~\ref{fig4}(a), we find that there
are obvious differences between them. First, the gain doublet is not opened near at $\omega=0$;
Second, the gain doublet is very narrow even for a very large control field. The physical reason
is that for the large control field the base state of the GAL-II system has very different
property from the GAN and GAL-I systems. In addition, there is no QI effect in the GAL-II system,
as shown below.

Fig.~\ref{fig6}(b) shows the group velocity of the signal field
 Re$(V_g)$ in the GAL-I system as a function of $\omega/\Gamma$ for
$\Omega_c=1.0\Gamma$. One sees that, as in the GAN and GAL-I systems, the GAL-II system can also
have subluminal and superluminal group velocities.

We can also make an analysis of the QI character in the GAL-II system
by using the SDM.
For $\Delta_3$ larger than $\gamma_{ij}$, $\Gamma_{ij}$, $\Omega_{p}$,
$\Omega_{\text{c}}$, Eq.~(\ref{K-II}) reduces to the form
$\tilde{K}(\omega)=\omega/c+\widetilde{\kappa}_{24}(\omega
+i\tilde{\gamma}_{41}/2)/[(\omega+i\gamma_{41})(\omega+i\gamma_{42}+|\Omega_{c}|^2/\Delta_3)]$,
which can be written as
\begin{equation}\label{K-IIS'}
K(\omega)=\frac{\omega}{c}+\widetilde{\kappa}_{24}\frac{\omega
+i\tilde{\gamma}_{41}/2}{(\omega-\omega_+)(\omega-\omega_-)},
\end{equation}
with $\widetilde{\kappa}_{24}=\kappa_{24}\sigma_{44}^{(0)}$,
$\tilde{\gamma}_{41}=\gamma_{41}+\gamma_{41}/(|\Omega_c|^2+1)$,
$\omega_{+}=-i\gamma_{41}$, and $\omega_{-}=-i\gamma_{42}-|\Omega_{c}|^2/\Delta_3$.
Equation (\ref{K-IIS'}) is different that for GAL-I system because now $\omega_{+}$ is
purely imaginary, which brings some different features for the
GAL-II system. First, the two peaks in the signal-field gain
spectrum $-$Im($K$) for non-zero $\Omega_{c}$ becomes asymmetric
(see the red dashed and black dotted-dashed lines in
Fig.~\ref{fig6}(a)\,) because the no real part exists for
$\omega_{+}$ but the real part of $\omega_-$ is proportional to
$|\Omega_{c}|^2$. This fact also explains why the small peak on the
right (the large peak on the left) of the gain dip locates at
$\omega=0$  (at $\omega=-|\Omega_{c}|^2/\Delta_3$). Second,
different from the GAN and GAL-I systems, in the GAL-II system
$\omega_-$ is always complex, so one cannot set up a parameter $\Omega_{\text{ref}}$
to divide the system into three control field regions.

It is easy to show that the gain spectrum of the GAL-II system can
be decomposed into
\begin{eqnarray}\label{form2-II}
-{\rm
Im}(K)=&&\frac{\widetilde{\kappa}_{24}}{2}\left\{\frac{f_1W_1}{\omega^2+W_1^2}
+\frac{f_2W_2}{(\omega+R_2)^2+W_2^2}\right.\nonumber\\
&&\left.\hspace{0.7cm}+\frac{g}{\delta}\left[\frac{\omega}{\omega^2+W_1^2}
-\frac{\omega+R_2}{(\omega+R_2)^2+W_2^2}\right]\right\},
\end{eqnarray}
with $W_1=\gamma_{41}$, $W_2=\gamma_{42}$, $R_2=|\Omega_c|^2/\Delta_3$,
$f_1=[-2W_1(W_2-W_1)+\tilde{\gamma}_{41}(W_2-W_1)]/\delta$,
$f_2=[2R_2^2+2W_2(W_2-W_1)-\tilde{\gamma}_{41}(W_2-W_1)]/\delta$,
$g=R_2(2W_1-\tilde{\gamma}_{41})$, and $\delta=R_2^2+(W_2-W_1)^2$.
The first two terms in the bracket of Eq.~(\ref{form2-II}) are
two Lorentzian terms. The following terms are proportional to $g/\delta$.
By a simple estimation one obtains
$g/\delta\simeq-\gamma_{41}/(2\Delta_3)$, which is negligibly small.
So the gain spectrum has only two Lorentzian terms and no interference
term, we thus conclude that there is no QI in GAL-II system and hence the
system displays only an ATS effect.

The GAL-II system has additional properties
absent in the GAN and GAL-I systems. For example,
it can becomes absorptive for a large control field. This
point has already shown in Fig.~\ref{fig6}(a), where the gain dip
opened in the black dotted-dashed line becomes deep enough so that
$-$Im($K$) becomes negative. The physical
reason is that for a large $\Omega_c$ the population
is mainly in the state $|4\rangle$,
resulting in a significant absorption of the
signal field.

Note that in addition to the level configuration indicated in the Sec.~\ref{Sec:3}
the GAL-I and GAL-II systems can also be realized by using the following
level configuration, i.e. $|1\rangle=|5^2S_{1/2},~F=1,~m_F=-1\rangle$,
$|2\rangle=|5^2S_{1/2},~F=2,~m_F=0\rangle$,
$|3\rangle=|5^2P_{1/2},~F=2,~m_F=0\rangle$, and
$|4\rangle=|5^2S_{3/2},~F=1,~m_F=0\rangle$.
In this case, the transition between $|1\rangle$ and  $|3\rangle$ is a single-photon
one, and the field coupled to $|2\rangle$ and $|4\rangle$ is a microwave or radiation-frequency
field~\cite{Zib}.

\section{Discussion and summary}\label{Sec:5}

From Sec.~\ref{Sec:2} to Sec.~\ref{Sec:4}, we have analyzed the QI
characters in the GAN, GAL-I, and GAL-II systems. For comparison, in
Table~\ref{Tab1}
\begin{table}
\caption{QI  characters of fast-light media}\label{Tab1}
\centering
\begin{tabular}{ccccc}
\hline    System & ~~Gain-doublet condition & ~~QI effect     & ~~ATS   \\
\hline    GAN    & ~~$|\Omega_c|^2 > \frac{\kappa_{23}|\Omega_p|^2\gamma_{41}}{\Delta^2_3}$
                 & ~~Constructive  & ~~Yes  \\
          GAL-I  & ~~$|\Omega_c|^2 > \frac{\kappa_{23}|\Omega_p|^2\gamma_{41}}{\Delta^2_3}$
                 & ~~Destructive   & ~~Yes  \\
          GAL-II & ~~$|\Omega_c|^2>\Delta_3\gamma_{42}$
                                 & ~~No            & ~~Yes  \\
\hline
\end{tabular}
\end{table}
we have summarized the main results obtained for different systems.
If in the table there is ``Yes'' in the same line for both QI and　
ATS, an QI-ATS crossover also exists in the system.

One may question possible influence resulted from
Doppler effect resulted from the thermal motion of the atoms,
which is omitted in the above discussions. When the
Doppler effect is considered, all calculations for the GAN, GAL-I
and GAL-II systems can still be carried out. As an example, in
Appendix~\ref{appA4} we present the result for the GAL-I system. The
conclusion is that, in the GAL-I system the QI and the crossover from
the QI to ATS are nearly the same as in the case without
Doppler effect if an experimental geometry of cancelling Doppler effect is adopted.

To show the difference between the three fast-light media with Doppler effect,
in Fig.~\ref{Fig7}
%
\begin{figure}
\centerline{\includegraphics[scale=1.0]{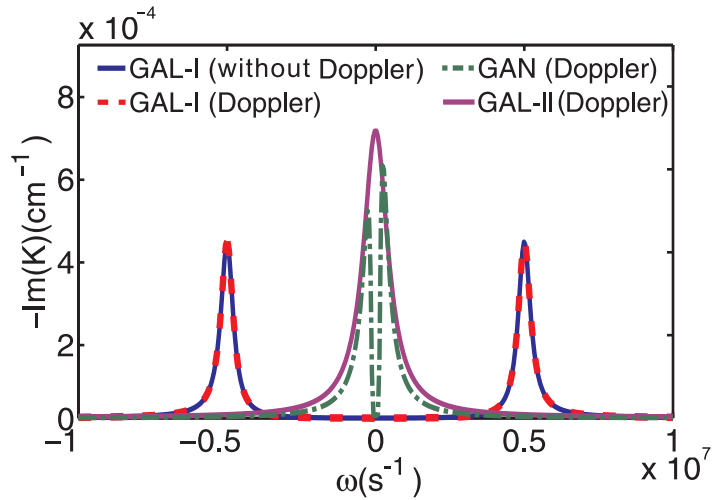}}
\caption{(Color online) Gain spectrum $-$Im($K$) as functions of $\omega$ for $\Omega_c=5.0\,\Gamma$.
The red dashed line (blue solid line) is for the GAL-I system with
(without) Doppler effect.
The green dashed-dotted line is for GAN system with Doppler effect.
The purple solid line is for the GAL-II system with Doppler effect,
which has been multiplied by $1/30$.}\label{Fig7}
\end{figure}
%
we show the gain spectrum $-$Im($K$) of the GAN, GAL-I and GAL-II
systems as functions of $\omega$ for $\Omega_c=5.0\,\Gamma$ when
Doppler broadening is taken into account. In the
figure, the red dashed line (blue solid line) is for the GAL-I
system with (without) Doppler effect; the green dashed-dotted line
is for GAN system with Doppler effect; the purple solid line is for
the GAL-II system with Doppler effect, which has been multiplied by
$1/30$ for display in the figure. We see that for the GAL-I system
there is almost no difference between the situations with and without
Doppler effect. However, it can be shown that for the GAN and the GAL-II systems, the QI
characters with Doppler effect have some differences in comparison with
the case without Doppler effect. Especially, when the Doppler effect is
present the width of the gain doublet for the GAN system becomes
very narrow and the two peaks become asymmetric and are amplified
(green dashed-dotted line).

In conclusion, in this article we have made a systematic
analysis on the QI effect in several fast-light media by using an extended
spectrum-decomposition method. We have shown that such fast-light
media are capable of not only completely eliminating the absorption
but also suppressing the gain of signal field, and hence provide the
possibility to realize a stable  long-distance propagation of the signal field with
a superluminal velocity. We have found that there is a destructive
(constructive) QI effect in GAL-I (GAN) system, but no QI
in the GAL-II system. We further found that a crossover from
destructive (constructive) QI to Autler-Townes splitting may happen for
the GAL-I (GAN) system when the control field of the system
is manipulated.

The fast-light media presented here may have giant Kerr nonlinearities,
as demonstrated in Refs.~\cite{Deng1,Hang1,Li,Tan,Zhu1,Zhu2}. The
fast, all-Optical, zero to $\pi$ continuously controllable phase gates
based on such media have been realized experimentally in Ref.~\cite{Deng2}.
The theoretical method presented here can be applied to other multi-level
atoms, or other physical systems (e.g. molecules, quantum dots,
nitrogen-valence centers in diamond, and rare-earth ions in crystals, etc.)
and the results obtained can help to deepen the understanding of fast-light physics
and fast-light spectroscopy and may have promising applications in optical
and quantum information processing and transmission, including the enhancement
of optical Kerr effect and the realization of light storage and retrieval by means of
fast light media~\cite{Aku}.

\acknowledgments

This work was supported by the NSF-China under Grants No.~11174080
and No.~11474099.

\appendix

\section{Bloch equations for the N system}\label{appA1}
The explicit expression of the Bloch Eq.~(\ref{BEGAN}) is given by
\begin{subequations}\label{Ntypea}
\begin{eqnarray}
&&i\frac{\partial}{\partial t}\sigma_{11}-i\Gamma_{13}\sigma_{33}-i\Gamma_{14}\sigma_{44}
+\Omega_{p}^{*}\sigma_{31}
-\Omega_{p}\sigma_{31}^{*}=0,\\
&&i\frac{\partial}{\partial t}\sigma_{22}-i\Gamma_{23}\sigma_{33}-i\Gamma_{24}
\sigma_{44}+\Omega_s^{*}\sigma_{32}
-\Omega_s\sigma_{32}^{*}+\Omega_{c}^*\sigma_{42}-\Omega_{c}\sigma_{42}^*=0,\\
&&i\left(\frac{\partial}{\partial
t}+\Gamma_{3}\right)\sigma_{33}+\Omega_{s}\sigma_{32}^{*}+\Omega_{p}\sigma_{31}^{*}
-\Omega_{s}^{*}\sigma_{32}-\Omega_{p}^{*}\sigma_{31}=0,\\
&&i\left(\frac{\partial}{\partial
t}+\Gamma_{4}\right)\sigma_{44}+\Omega_{c}\sigma_{42}^{*}-\Omega_{c}^{*}\sigma_{42}=0,
\end{eqnarray}
\end{subequations}
for the diagonal matrix elements, and
\begin{subequations}\label{Ntypeb}
\begin{eqnarray}
&&\left(i\frac{\partial}{\partial
t}+d_{21}\right)\sigma_{21}+\Omega_{c}^{*}\sigma_{41}
+\Omega_s^*\sigma_{31}-\Omega_{p}\sigma_{32}^{*}=0,\\
&&\left(i\frac{\partial}{\partial
t}+d_{31}\right)\sigma_{31}+\Omega_{p}(\sigma_{11}
-\sigma_{33})+\Omega_{s}\sigma_{21}=0,\\
&&\left(i\frac{\partial}{\partial
t}+d_{32}\right)\sigma_{32}+\Omega_{p}\sigma_{21}^{*}
+\Omega_{s}(\sigma_{22}-\sigma_{33})-\Omega_{c}\sigma_{43}^*=0,\\
&&\left(i\frac{\partial}{\partial
t}+d_{41}\right)\sigma_{41}+\Omega_{c}\sigma_{21}
-\Omega_{p}\sigma_{43}=0,\\
&&\left(i\frac{\partial}{\partial
t}+d_{42}\right)\sigma_{42}+\Omega_{c}(\sigma_{22}
-\sigma_{44})-\Omega_s\sigma_{43}=0,\\
&&\left(i\frac{\partial}{\partial
t}+d_{43}\right)\sigma_{43}+\Omega_{c}\sigma_{32}^{*}
-\Omega_{p}^*\sigma_{41}-\Omega_s^*\sigma_{42}=0,
\end{eqnarray}
\end{subequations}
for the off-diagonal matrix elements, where $d_{21}=\Delta_{2}+i\gamma_{21}$,
$d_{31}=\Delta_{3}+i\gamma_{31}$, $d_{32}=\Delta_{3}-\Delta_{2}+i\gamma_{32}$,
$d_{41}=\Delta_{4}+i\gamma_{41}$, $d_{42}=\Delta_{4}-\Delta_{2}+i\gamma_{42}$,
$d_{43}=\Delta_{4}-\Delta_{3}+i\gamma_{43}$ with
$\gamma_{jl}=(\Gamma_{j}+\Gamma_{l})/2+\gamma_{jl}^{\rm dep}$. Here
$\Gamma_{j}=\sum_{l<j}\Gamma_{lj}$ with $\Gamma_{jl}$ ($\gamma_{jl}^{\rm dep}$) the population
decay rate (dephasing rate) from the level $|l\rangle$ and $|j\rangle$ ($j,l=1$-4).

\section{Bloch equations for the GAL-I system}\label{appA2}

The Bloch equations of the GAL-I system read
\begin{subequations}\label{Ladder-Ia}
\begin{eqnarray}
&&i\frac{\partial}{\partial
t}\sigma_{11}-i\Gamma_{13}\sigma_{33}-i\Gamma_{14}\sigma_{44}+\Omega_{p}^{*}\sigma_{31}
-\Omega_{p}\sigma_{31}^{*}=0,\\
&&i\left(\frac{\partial}{\partial
t}+\Gamma_{42}\right)\sigma_{22}-i\Gamma_{23}\sigma_{33}+\Omega_s^{*}\sigma_{32}
-\Omega_s\sigma_{32}^{*}+\Omega_{c}^*\sigma_{42}-\Omega_{c}\sigma_{42}^*=0,\\
&&i\left(\frac{\partial}{\partial
t}+\Gamma_{3}\right)\sigma_{33}+\Omega_{s}\sigma_{32}^{*}+\Omega_{p}\sigma_{31}^{*}
-\Omega_{s}^{*}\sigma_{32}-\Omega_{p}^{*}\sigma_{31}=0,\\
&&i\left(\frac{\partial}{\partial
t}+\Gamma_{4}\right)\sigma_{44}-i\Gamma_{42}\sigma_{22}+\Omega_{c}\sigma_{42}^{*}
-\Omega_{c}^{*}\sigma_{42}=0,
\end{eqnarray}
\end{subequations}
for the off-diagonal matrix elements, and
\begin{subequations}\label{Ladder-Ib}
\begin{eqnarray}
&&\left(i\frac{\partial}{\partial
t}+d_{21}\right)\sigma_{21}+\Omega_{c}^{*}\sigma_{41}+\Omega_s^*\sigma_{31}-\Omega_{p}\sigma_{32}^{*}=0,\\
&&\left(i\frac{\partial}{\partial
t}+d_{31}\right)\sigma_{31}+\Omega_{p}(\sigma_{11}-\sigma_{33})+\Omega_{s}\sigma_{21}=0,\\
&&\left(i\frac{\partial}{\partial
t}+d_{32}\right)\sigma_{32}+\Omega_{p}\sigma_{21}^{*}+\Omega_{s}(\sigma_{22}-\sigma_{33})-\Omega_{c}\sigma_{43}^*=0,\\
&&\left(i\frac{\partial}{\partial
t}+d_{41}\right)\sigma_{41}+\Omega_{c}\sigma_{21}-\Omega_{p}\sigma_{43}=0,\\
&&\left(i\frac{\partial}{\partial
t}+d_{42}\right)\sigma_{42}+\Omega_{c}(\sigma_{22}-\sigma_{44})-\Omega_s\sigma_{43}=0,\\
&&\left(i\frac{\partial}{\partial
t}+d_{43}\right)\sigma_{43}+\Omega_{c}\sigma_{32}^{*}-\Omega_{p}^*\sigma_{41}-\Omega_s^*\sigma_{42}=0,
\end{eqnarray}
\end{subequations}
for the off-diagonal matrix elements,
where $d_{21}=\Delta_{2}-\Delta_{1}+i\gamma_{21}$,
$d_{31}=\Delta_{3}-\Delta_{1}+i\gamma_{31}$,
$d_{32}=\Delta_{3}-\Delta_{2}+i\gamma_{32}$,
$d_{41}=\Delta_{4}-\Delta_{1}+i\gamma_{41}$,
$d_{42}=\Delta_{4}-\Delta_{2}+i\gamma_{42}$, and
$d_{43}=\Delta_{4}-\Delta_{3}+i\gamma_{43}$.

The base state of the GAL-I system (i.e. the steady-state solution
of the MB Eqs.~(\ref{Ladder-Ia}), (\ref{Ladder-Ib}), and
(\ref{Maxwell-N2}) for  $\Omega_s=0$) is given by
\begin{subequations} \label{BS2}
\begin{eqnarray}
&&\sigma_{11}^{(0)}=\frac{\Gamma_{14}(|\Omega_{c}|^2+\Gamma_{42}X_{42})(\Gamma_3X_{31}
+|\Omega_{p}|^2)}{D},\\
&&\sigma_{22}^{(0)}=\frac{\Gamma_{23}|\Omega_{p}|^2(\Gamma_4X_{42}+|\Omega_{c}|^2)}{D},\\
&&\sigma_{33}^{(0)}=\frac{\Gamma_{14}(|\Omega_{c}|^2+\Gamma_{42}X_{42})|\Omega_{p}|^2}{D},\\
&&\sigma_{44}^{(0)}=\frac{\Gamma_{23}(|\Omega_{c}|^2+\Gamma_{42}X_{42})|\Omega_{p}|^2}{D},\\
&&\sigma_{31}^{(0)}=-\frac{\Omega_{p}}{d_{31}}\frac{\Gamma_{14}\Gamma_3X_{31}(|\Omega_{c}|^2
+\Gamma_{42}X_{42})}{D},\\
&&\sigma_{42}^{(0)}=-\frac{\Omega_{c}}{d_{42}}\frac{\Gamma_{23}(\Gamma_4
-\Gamma_{42})X_{42}|\Omega_{p}|^2}{D},
\end{eqnarray}
\end{subequations}
and other $\sigma_{jl}^{(0)}=0$. Here we have defined $X_{31}=|d_{31}|^2/(2\gamma_{31})$,
$X_{42}=|d_{42}|^2/(2\gamma_{42})$ and
$D=\Gamma_{14}(|\Omega_{c}|^2+\Gamma_{42}X_{42})(\Gamma_3X_{31}+2|\Omega_{p}|^2)
+\Gamma_{23}|\Omega_{p}|^2[(\Gamma_4-\Gamma_{42})X_{42}+2(|\Omega_{c}|^2+\Gamma_{42}X_{42})]$.
For large $\Delta_3$, the base state solution is simplified into
$\sigma_{11}^{(0)}\approx 1$, $\sigma_{31}^{(0)}\approx -\Omega_p/d_{31}$,
and all other $\sigma_{jl}^{(0)}\approx 0$.


\section{Base state solution of the GAL-II system}\label{appA3}

The base state solution of the GAL-II system is

\begin{subequations} \label{BS2'}
\begin{eqnarray}
&&\sigma_{22}^{(0)}=\frac{\alpha_{12}\alpha_{23}-\alpha_{22}\alpha_{13}}
{\alpha_{11}\alpha_{22}-\alpha_{12}\alpha_{21}},\\
&&\sigma_{33}^{(0)}=\frac{\alpha_{13}\alpha_{21}-\alpha_{11}\alpha_{23}}
{\alpha_{11}\alpha_{22}-\alpha_{12}\alpha_{21}},\\
&&\sigma_{44}^{(0)}=\frac{\Gamma_{42}}{\Gamma_4}\sigma_{22}^{(0)},\\
&&\sigma_{11}^{(0)}=1-\sigma_{22}^{(0)}-\sigma_{33}^{(0)}-\sigma_{44}^{(0)},\\
&&\sigma_{21}^{(0)}=\frac{\Omega_{p}\Omega^*_{\text{c}}\left[d^*_{32}-\left(\frac{\Gamma_{42}}{\Gamma_4}
d^*_{32}+d_{31}\right)\sigma_{22}^{(0)}+(d_{31}-2d^*_{32})\sigma_{33}^{(0)}\right]}{B},\\
&&\sigma_{31}^{(0)}=\frac{-\Omega_{\text{c}}\sigma_{21}^{(0)}-\Omega_{p}(\sigma_{11}^{(0)}-\sigma_{33}^{(0)})}{d_{31}},\\
&&\sigma_{32}^{(0)}=\frac{-\Omega_{p}\sigma_{21}^{(0)*}-\Omega_{\text{c}}(\sigma_{22}^{(0)}-\sigma_{33}^{(0)})}{d_{32}},
\end{eqnarray}
\end{subequations}
with
\begin{subequations} \label{BSA}
\begin{eqnarray}
& & B = d_{21}d^*_{32}d_{31}+d_{31}|\Omega_{p}|^2-d^*_{32}|\Omega_{\text{c}}|^2, \\
& & \alpha_{11}=-i\Gamma_{42}+|\Omega_{p}|^2|\Omega_{\text{c}}|^2
\left(1+\frac{\Gamma_{42}}{\Gamma_4}\right)\left(\frac{d^*_{32}}{Bd_{31}}-
\frac{d_{32}}{B^*d^*_{31}}\right)\nonumber\\
& & \hspace{1.2cm}+|\Omega_{p}|^2|\Omega_{\text{c}}|^2\left(\frac{1}{B}-\frac{1}{B^*}\right)+
|\Omega_{p}|^2\left(1+\frac{\Gamma_{42}}{\Gamma_4}\right)
\left(\frac{2i\gamma_{31}}{|d_{31}|^2}\right),\\
& &
\alpha_{12}=-i\Gamma_{13}+2|\Omega_{p}|^2\left(\frac{1}{d_{31}}-\frac{1}{d^*_{31}}\right)+
2|\Omega_{p}|^2|\Omega_{\text{c}}|^2\left(1+\frac{\Gamma_{42}}{\Gamma_4}\right)
\left(\frac{d^*_{32}}{Bd_{31}}
-\frac{d_{32}}{B^*d^*_{31}}\right)\nonumber\\
& & \hspace{1.2cm}+|\Omega_{p}|^2|\Omega_{\text{c}}|^2\left(\frac{1}{B^*}-\frac{1}{B}\right),\\
& & \alpha_{13}=|\Omega_{p}|^2|\Omega_{\text{c}}|^2\left(\frac{d_{32}}{B^*d^*_{31}}
-\frac{d^*_{32}}{Bd_{31}}\right)+
|\Omega_{p}|^2\left(\frac{1}{d^*_{31}}-\frac{1}{d_{31}}\right),
\end{eqnarray}
\end{subequations}
and
\begin{subequations} \label{BSA1}
\begin{eqnarray}
& & \alpha_{21}=i\Gamma_{42}+|\Omega_{p}|^2|\Omega_{\text{c}}|^2
\left(1+\frac{\Gamma_{42}}{\Gamma_4}\right)\left(\frac{1}{B^*}-\frac{1}{B}\right)+
|\Omega_{p}|^2|\Omega_{\text{c}}|^2\left(\frac{d^*_{31}}{B^*d_{32}}-\frac{d_{31}}{Bd^*_{32}}
\right)\nonumber\\
& & \hspace{1.2cm}+
|\Omega_{\text{c}}|^2\left(\frac{1}{d^*_{32}}-\frac{1}{d_{32}}\right),\\
& &
\alpha_{22}=-i\Gamma_{23}+2|\Omega_{p}|^2|\Omega_{\text{c}}|^2\left(\frac{1}{B^*}-\frac{1}{B}\right)+
|\Omega_{p}|^2|\Omega_{\text{c}}|^2\left(\frac{d_{31}}{Bd^*_{32}}
-\frac{d^*_{31}}{B^*d_{32}}\right)\nonumber\\
& & \hspace{1.2cm}+
|\Omega_{\text{c}}|^2\left(\frac{1}{d_{32}}-\frac{1}{d^*_{32}}\right),\\
& &
\alpha_{23}=|\Omega_{p}|^2|\Omega_{\text{c}}|^2\left(\frac{1}{B}-\frac{1}{B^*}\right).
\end{eqnarray}
\end{subequations}
%

\section{QI effect in the GAL-I system with Doppler effect}\label{appA4}

When the Doppler effect is considered, the linear dispersion relation of the GAL-I system
reads
\begin{equation}\label{K-IS-Dopp1}
K(\omega)=\frac{\omega}{c}+\widetilde{\kappa}_{23}\int^{\infty}_{-\infty}f(v)\frac{\omega+d_{41}}{(\omega+d_{21})
(\omega+d_{41})-|\Omega_{c}|^2}dv
\end{equation}
with
$d_{41}=\Delta_4+i\gamma_{41}-(k_p-k_c-k_s)v$, $d_{21}=\Delta_2+i\gamma_{21}-(k_p-k_s)v$,
$f(v)$ is Maxwell velocity distribution function. To show clearly whether the QI
still keeps or not, as done in Refs.~~\cite{Kuznetsova2002,Lee2003,Li2010}
we replace the Maxwell velocity distribution function
by a modified Lorentzian velocity distribution $f(v)=v_T/[\sqrt{\pi}(v^2+v^2_T)]$, where
$v_T=\sqrt{2k_B T/M}$ is the most probable speed at temperature $T$
with $k_B$ the Boltzmann constant and $M$ the particle mass.
We assume that all three light fields are incident in the same (i.e. $z$)
direction (i.e. an ``almost Doppler-free'' geometry~\cite{XiaoM}).
Since in this situation $k_p-k_s-k_s\approx 0$,
one has $d_{41} \approx \Delta_4+i\gamma_{41}$ and hence
the integrand in the integral of Eq.~(\ref{K-IS-Dopp1}) can be simplified largely.
By using the residue theorem~\cite{Kuznetsova2002,Lee2003,Li2010},
one can carry out the integral  analytically, yielding the result
$K(\omega)=\omega/c+\widetilde{\kappa}_{23}\sqrt{\pi}(\omega+i\gamma_{41})/[(\omega+i\gamma_{21}+i\Delta\omega_D)
(\omega+i\gamma_{41})-|\Omega_{c}|^2]$ ($\Delta\omega_D=k_s v_T$ is Doppler width), which can be rewritten as
\begin{equation}\label{K-dI-Dopp}
K(\omega)=\frac{\omega}{c}+\widetilde{\kappa}_{23}\frac{\sqrt{\pi}(\omega
+i\gamma_{41})}{(\omega-\omega_+)(\omega-\omega_-)},
\end{equation}
with
$\omega_{\pm}=-i(\gamma_{21}+\gamma_{41}+\Delta\omega_D)/2
\pm \sqrt{\left[|\Omega_{c}|^2-|\Omega_{\rm ref}|^2\right]}$
and $\Omega_{\rm ref}=|\gamma_{21}+\Delta\omega_D-\gamma_{41}|/2$.
Equation (\ref{K-dI-Dopp}) is similar to Eq.~(\ref{K-dI}), with
difference only in $\omega_{\pm}$ and $\Omega_{\rm ref}$ which
are now dependent on $\Delta\omega_D$ due to
the Doppler effect.

From Eq.~(\ref{K-dI-Dopp}) one can obtain the gain spectrum $-\text{Im}(K)$, which
can be decomposed  in different control-field regions by using the SDM.
Similar formulas (like (\ref{Kd2})-(\ref{form2-I})\,) and figure (like
Fig.~\ref{fig5}) can be obtained, which are omitted here. From these results we can acquire the
following conclusions:
(i)~In the weak control field region,  $-\text{Im}(K)$ is the sum of
two Lorentzian terms centered at $\omega=0$, which have opposite signs.
The superposition of the two Lorentzian terms results in a quantum destructive interference
and hence a gain doublet appears in the gain spectrum. (ii)~In the
strong control field region,  $-\text{Im}(K)$ can be approximately
expressed as a sum of two Lorentzian terms, which however have the same sign, locate
at different positions, and far apart each other. Thus in this region the gain spectrum is an ATS one because
there is no interference occurring.
(iii)~In the intermediate control field region, which is the region between the weak and the strong ones,
the gain spectrum displays a crossover from the quantum destructive interference to the ATS.

In a similar way, one can make similar calculations for the GAN and GAL-II systems.
The general conclusions obtained in Sec.~III and Sec.~IV are not changed when Doppler effect
is taken into consideration. However, QI characters in the GAN and GAL-II systems with Doppler effect
display differences in comparison with that without Doppler effect, some of which have been
described in Fig.~\ref{Fig7}.



\begin{thebibliography}{}

\bibitem{Milo05}P. W. Milonni, {\it Fast Light, Slow Light and Left-handed Light}  (Inst. Phys.
    Pub., Bristol, 2005).

\bibitem{Khurgin2009} K. B. Khurgin and R. S. Tucker (ed), {\it Slow Light: Science and
    Applications} (Boca Raton, Taylor and Francis, 2009).

\bibitem{Fleischhauer2005} M. Fleischhauer, A. Imamoglu, and J. P. Marangos, ``Electromagnetically
    induced transpa rency: Optics in coherent media,'' Rev. Mod. Phys. {\bf 77}, 633 (2005).

\bibitem{Deng1} L. Deng, M. G. Payne, ``Gain-assisted large and rapidly responding Kerr effect
    using a room-temperature active Raman gain medium,'' Phys. Rev. Lett. {\bf 98}, 253902 (2007).

\bibitem{note00}Slow light (fast light) refers to the situation in which the group velocity of a
    light pulse $V_g<<c$ ($V_g>c$ or negative)~\cite{Milo05}.

\bibitem{Boyd1}R. W. Boyd and D. J. Gauthier, ``Slow and Fast Light,'' {\it Progress in Optics}
    (Elsevier Science, 2002), Vol. 43, Chap. 6, p. 275  and references therein.

\bibitem{Boyd2}R. W. Boyd and D. J. Gauthier, ``Controlling the velocity of light pulses,''
    Science {\bf 326}, 1704 (2009).

\bibitem{Chu}S. Chu and S. Wong, ``Linear pulse propagation in an absorpting medium,'' Phys. Rev.
    Lett. {\bf 48}, 738 (1982).

\bibitem{Chiao}R. Y. Chiao, ``Superluminal (but causal) propagation of wave packets in transparent
    media with inverted atomic populations,'' Phys. Rev. A {\bf 48}, R34 (1993).

\bibitem{Steinberg}A. M. Steinberg and R. Y. Chiao, ``Dispersionless, highly superluminal
    propagation in a medium with a gain doublet,''  Phys. Rev. A {\bf 49}, 2071 (1994).

\bibitem{Wang} L. J. Wang, A. Kuzmich, A. Dogariu, ``Gain-assisted superluminal light
    propagation,'' Nature {\bf 406}, 277 (2000).

\bibitem{Bigelow} M. S. Bigelow, N. N. Lepeshkin, R. W. Boyd, ``Superluminal and slow light
    propagation in a room-temperature solid,'' Science {\bf 301}, 200 (2003).

\bibitem{Agarwal2004} G. S. Agarwal, S. Dasgupta, ``Superluminal propagation via coherent
    manipulation of the Raman gain process,'' Phys. Rev. A. {\bf 70}, 023802 (2004).

\bibitem{Jiang1} K. J. Jiang, L. Deng, and M. G. Payne, ``Ultraslow propagation of an optical
    pulse in a three-state active Raman gain medium,'' Phys. Rev. A {\bf 74}, 041803 (2006).

\bibitem{Jiang2} K. J. Jiang, L. Deng, and M. G. Payne, ``Superluminal propagation of an optical
    pulse in a Doppler-broadened three-state single-channel active Raman gain medium,'' Phys. Rev.
    A {\bf 76}, 033819 (2007).

\bibitem{Huang1}G. Huang, C. Hang, and L. Deng, ``Gain-assisted superluminal optical solitons at
    very low light intensity,'' Phys. Rev. A {\bf 77}, 011803(R) (2008).

\bibitem{Hang1} C. Hang, and G. Huang, ``Giant Kerr nonlinearity and weak-light superluminal
    optical solitons in a four-state atomic system with gain doublet,'' Opt. Express {\bf 18},
    2954 (2010).

\bibitem{Li} H. Li, L. Dong, C. Hang, and G. Huang, ``Gain-assisted high-dimensional self-trapped
    laser beams at very low light levels,'' Phys. Rev. A {\bf 83}, 023816 (2011).

\bibitem{Zhu1}C. J. Zhu, C. Hang, and G. Huang, ``Gain-assisted giant Kerr nonlinearity in a
    $\Lambda$-type system with two-folded lower levels,'' Eur. Phys. J. D {\bf 56}, 231 (2010).

\bibitem{Zhu2} C. J. Zhu, and G. Huang, ``High-order nonlinear Schr\"{o}dinger equation and
    weak-light superluminal solitons in active Raman gain media with two control fields,'' Opt.
    Express {\bf 19}, 1963 (2011).

\bibitem{Deng2}R. B. Li, L. Deng, and E. W. Hagley, ``Fast, All-Optical, Zero to $\pi$
    Continuously Controllable Kerr Phase Gate,'' Phys. Rev. Lett. {\bf 110}, 113902 (2013).

\bibitem{Tan}C. Tan and G. Huang, ``Surface polaritons in a negative-index metamaterial with
    active Raman gain,'' Phys. Rev. {\bf 91}, 023803 (2015).

\bibitem{Hang2}C. Hang and G. Huang, ``Highly entangled photons and rapidly responding
    polarization qubit phase gates in a room-temperature active Raman gain medium,'' Phys. Rew. A
    {\bf 82}, 053818 (2010).

\bibitem{Lezama} A. Lezama, A. M. Akulshin, A. I. Sidorov, P. Hannaford, ``Storage and retrieval
    of light pulses in atomic media with ``slow'' and ``fast'' light,'' Phys. Rev. A. {\bf 73},
    033806 (2006).

\bibitem{Lett}J. B. Clark, R. T. Glasser, Q. Glorieux, U. Vogl, T. Li, K. M. Jones, and P. D.
    Lett, ``Quantum mutual information of an entangled state propagating through a fast-light
    medium,'' Nat. Photon. {\bf 8}, 515 (2014).

\bibitem{Agarwal1997} G. S. Agarwal, ``Nature of the quantum interference in
    electromagnetic-field-induced control of absorption,'' Phys. Rev. A {\bf 55}, 2467 (1997).

\bibitem{Abi-Salloum2010} T. Y. Abi-Salloum, ``Electromagnetically induced transparency and
    Autler-Townes splitting: two similar but distinct phenomena in two categories of three-level
    atomic systems,'' Phys. Rev. A {\bf 81}, 053836 (2010).

\bibitem{Anisimov2011} P. M. Anisimov, J. P. Dowling, and B. C. Sanders, ``Objectively discerning
    Autler-Townes splitting from electromagnetically induced transparency,'' Phys. Rev. Lett {\bf
    107}, 163604 (2011).

\bibitem{Giner2013} L. Giner, L. Veissier, B. Sparkes, A. S. Sheremet, A. Nicolas, O. S. Mishina,
    M. Scherman, S. Burks, I. Shomroni, D. V. Kupriyanov, P. K. Lam, E. Giacobino, and J. Laurat,
    ``Experimental investigation of the transition between Autler-Townes splitting and
    electromagnetically-induced-transparency models,'' Phys. Rev. A {\bf 87}, 013823 (2013).

\bibitem{Tan2013} C. Tan, C. Zhu, and G. Huang, ``Analytical approach on linear and nonlinear
    pulse propagations in an open $\Lambda$-type molecular system with Doppler broadening,'' J.
    Phys. B {\bf 46}, 025103 (2013).

\bibitem{Zhu2013} C. Zhu, C. Tan, and G. Huang, ``Crossover from electromagnetically induced
    transparency to Autler-Townes splitting in open V-type molecular systems,'' Phys. Rev. A {\bf
    87}, 043813 (2013).

\bibitem{Tan2014} C. Tan and G. Huang, ``Crossover from electromagnetically induced transparency
    to Autler-Townes splitting in open ladder systems with Doppler broadening,'' J. Opt. Soc. Am.
    B {\bf 31}, 704 (2014).

\bibitem{Peng2014} Peng. B, S. K. Ozdemir, W. J. Chen,  F. Nori, L. Yang, ``What is and what is
    not electromagnetically induced transparency in whispering-gallery microcavities,'' Nat.
    Commun. {\bf 5}, 5085 (2014).

\bibitem{Ding}J. Ding, B. Arigong, H. Ren, M. Zhou, J. Shao, M. Lu, Y. Chai, Y. Lin, and H. Zhang,
    ``Tuneable complementary metamaterial structures based on graphene for single and multiple
    transparency windows,'' Sci. Rep. {\bf 4}, 6128 (2014).

\bibitem{Lu}X. Lu, X. Miao, J. Bai, L. Pei, M. Wang, Y. Gao, L.-A. Wu, P. Fu, R. Wang, and Z. Zuo,
    ``Transition from Autler-Townes splitting to electromagnetically induced transparency based on
    the dynamics of decaying dressed states,'' J. Phys. B: At. Mol. Opt. Phys. {\bf 48}, 055003
    (2015).

\bibitem{Davuluri}S. Davuluria, Y. Wang and S. Zhu, ``Destructive and constructive interference in
    the coherently driven three-level systems,''  J. Mod. Opt. {\bf 62}, 1091 (2015).

\bibitem{Boyd}R. W. Boyd, {\it Nonlinear Optics} (3rd edition) (Academic, Elsevier, 2008).

\bibitem{note000}Through the text, all signal fields in the GAN, GAL-I and GAL-II systems are
    assumed to be weak so that their nonlinear effect is negligible.

\bibitem{note1}The frequency and wave number of the signal field are given by $\omega_s+\omega$
    and $k_s+K(\omega)$, respectively. Thus $\omega=0$ corresponds to the center frequency of the
    signal field.

\bibitem{Zib}A. S. Zibrov, A. B. Matsko, and M. O. Scully, ``Four-Wave Mixing of Optical and
    Microwave Fields,'' Phys. Rev. Lett. {\bf 89}, 103601 (2002).

\bibitem{Kuznetsova2002}E. Kuznetsova, O. Kocharovskaya, P. Hemmer, and M. O. Scully, ``Atomic
    interference phenomena in solids with a long-lived spin coherence,'' Phys. Rev. A {\bf 66},
    063802 (2002).

\bibitem{Lee2003}H. Lee, Y. Rostovtsev, C. J. Bednar, and A. Javan, ``From laser-induced line
    narrowing to electromagnetically induced transparency: closed system analysis,'' Appl. Phys. B
    {\bf 76}, 33 (2003).

\bibitem{Li2010}L. Li and G. Huang, ``Linear and nonlinear light propagations in a
    Doppler-broadened medium via electromagnetically induced transparency,'' Phys. Rev. A {\bf
    82}, 023809 (2010).

\bibitem{XiaoM}M. Xiao, Y.-q. Li, S.-z. Jin, and J. Gea-Banacloche, ``Measurement of Dispersive
    Properties of Electromagnetically Induced Transparency in Rubidium Atoms,'' Phys. Rev. Lett.
    {\bf 74}, 666 (1995).

\bibitem{Aku}A. M.  Akulshin and R. J. McLean, ``Fast light in atomic media,'' J. Opt. {\bf 12},
    104001 (2010).

\end{thebibliography}
\end{document}